\documentclass{article}
\usepackage{arxiv}
\usepackage[utf8]{inputenc} 
\usepackage{hyperref}       
\usepackage{amsfonts}       
\usepackage{amsmath}
\usepackage{color}
\usepackage{graphicx}
\usepackage{subfigure}
\usepackage{cite}
\usepackage{braket}
\usepackage[title]{appendix} 

\title{Phase-space representation of coherent states generated through SUSY QM for tilted anisotropic Dirac materials}
\author{Daniel O-Campa$^{1}$ and Erik Díaz-Bautista$^{2}$\\
Escuela Superior de Fisica y Matematicas del Instituto Politecnico Nacional, 07738 Mexico City, Mexico.\\
e-mail: dortizca@ipn.mx$^1$, ediazba@ipn.mx$^2$}
\begin{document}
\maketitle
\begin{abstract}
In this paper, we examine the electron interaction within tilted anisotropic Dirac materials when subjected to external electric and magnetic fields possessing translational symmetry. Specifically, we focus on a distinct non-zero electric field magnitude, enabling the decoupling of the differential equation system inherent in the eigenvalue problem. Subsequently, employing supersymmetric quantum mechanics facilitates the determination of eigenstates and eigenvalues corresponding to the Hamiltonian operator. To delve into a semi-classical analysis of the system, we identify a set of coherent states. Finally, we assess the characteristics of these states using fidelity and the phase-space representation through the Wigner function.
\end{abstract}
\section{Introduction}
In condensed matter area, anisotropic Dirac materials with tilted cones have been extensively studied because of their unique electronic properties and potential applications in various fields of physics \cite{Zhao2018,Feng20161, Mannix2015, LopezBezanilla2016,Goerbig2009, katsnelson_2012,Cheng2017}. 
A main characteristic of these materials is the behavior of their charge carriers near the Dirac points is described by an effective Hamiltonian that also incorporates both anisotropy and the tilting of the Dirac cones. This fact has led to an emergent research area in semiconductor technology, namely valleytronics \cite{Schaibley2016, Ang2017}, in which the valley degree of freedom could allow the manipulation of the electronic transport in two-dimensional materials, even in pristine graphene, in which the inversion symmetry of the system prohibits valley-selective excitations \cite{Mrudul21}. \\
\\
Thereby, to deepen the study of such materials, coherent states, which arose from the study of harmonic oscillators, have been extended to the analysis of interactions of Dirac materials with external fields, leading to the formulation of so-called Barut-Girardello coherent states among others \cite{DiazBautista2020, diaz2022time, betancur2021}, which play a crucial role in the investigation on phase space. Besides, supersymmetric quantum mechanics (SUSY QM) approach can be used to solve the eigenvalue problem associated with the effective Hamiltonian and allows the construction of annihilation operators \cite{Kuru2009,Midya_2014,Fernández_C_2020,Fernández_C_2021}.  In a complementary way, the fidelity between an evolved state and its initial state provides insights into the periodic behavior of the system \cite{FernándezC.2022}. Likewise, the Wigner function \cite{Wigner1932}, a quasiprobability distribution, offers a perspective on the distribution of coherent states in phase space, shedding light on their classical or quantum nature \cite{Case2008, 1977a, Kenfack2004, Smithey1993}. Initially, the Wigner function arose from a quantum mechanics formulation in phase space \cite{Hillery1984, Zachos2005} in which, under certain requirements, it is possible to associate an integrable function defined on $\mathbb{R}^{2n}$ to an operator in a Hilbert space $\mathcal{H}$ through the so-called Weyl transform \cite{BAYEN1978I, BAYEN1978II, Bordemann_2008}. Nowadays, there are many experimental applications of the Wigner function in quantum transport studies, since it allows a mixed quantum--semi-classical description of the systems \cite{Weinbub2018, Weinbub_2022}. For instance, the Wigner function has been employed in a quantum-tomography approach to reconstruct quantum states of solitary electrons or electric currents \cite{Leibfried1998, Andrea_Bertoni_1999, Carlo_Jacoboni_2004, Jullien2014, Bisognin2019, Fletcher2019, Roussel2021}. \\
\\
With this motivation, this paper presents a systematic exploration of anisotropic Dirac materials with tilted cones as follows: in section \ref{1} it is introduced the application of SUSY QM to determine eigenstates and eigenvalues of the Hamiltonian operator {\color{black}assuming a specific electric field amplitude that enables such a task}. Then, in section \ref{2} the concept of Barut-Girardello coherent states is discussed. After that,  in section \ref{3} the behavior of the Barut-Girardello coherent states is discussed by using fidelity and the Wigner function. Finally, in section \ref{4} we discuss the conclusions. 
\section{The eigenvalue problem}\label{1}
The effective Hamiltonian that describes the charge carriers at low energies in Dirac-like materials with anisotropy and tilted cones is given by 
\begin{equation}
\mathcal{H}_0 = \nu \left( v_x p_x \sigma_x + v_y p_y \sigma_y + v_t p_y \sigma_0 \right),
\label{eq:1}
\end{equation}
in this scenario, $\nu = \pm 1$ denotes the valley index, $v_x$ and $v_y$ stand for the anisotropic velocities, and $v_t$ represents a velocity component originating from the tilting of the Dirac cones. In this context, $\sigma_{x},\sigma_{y}$ represent the Pauli matrices, $\sigma_0$ is the $2\times2$ identity matrix, and $p_x$, $p_y$ denote the canonical momentum operators. When we account for the impact of magnetic and electric fields that are stationary but vary with position, the Hamiltonian described in Equation (\ref{eq:1}) needs adjustment following the minimal coupling rule. Consequently, in a simplified scenario where both fields vary along a single axis (specifically the $x$-axis), and the magnetic field $\vec{\mathcal{B}}$ is perpendicular to the surface of the material (the $x$-$y$ plane), with the electric field $\vec{\mathcal{E}}$ confined within the plane, the Hamiltonian transforms into
\begin{equation}
\mathcal{H} = \nu v_x \left[ p_x \sigma_x + \frac{v_y}{v_x} \left(p_y + \mathcal{A}_y(x)\right)\sigma_y + \frac{v_t}{v_x} \left(p_y + \mathcal{A}_y(x) - \frac{\nu}{v_t}\phi(x)\right) \sigma_0 \right],
\label{eq:2}
\end{equation}
where $\vec{\mathcal{A}} = \mathcal{A}_y(x)\hat{e}_y$ represents the vector potential, leading to $\vec{\mathcal{B}}=\nabla\times\vec{\mathcal{A}}=\mathcal{B}(x)\hat{e}_z$ ($\mathcal{B}(x)=\mathcal{A}'_y(x)$), and $\phi(x)$ denotes the scalar potential, yielding $\vec{\mathcal{E}}=-\nabla\phi(x)=\mathcal{E}(x)\hat{e}_x$ ($\mathcal{E}(x)=-\phi'(x)$). It is evident from this Hamiltonian that $\left[\mathcal{H},p_y\right]=0$. Consequently, in the eigenvalue equation $\mathcal{H}\Psi(x,y) = E\Psi(x,y)$, the eigenfunctions can be represented as
\begin{equation}
\Psi(x,y) = \mathrm{e}^{iky}\bar{\Psi}(x),
\label{eq:3}
\end{equation}
where $\bar{\Psi}(x) = \begin{pmatrix} \psi^+(x), i\psi^-(x) \end{pmatrix}^T$.
Then, the Hamiltonian acting on $\Psi(x,y)$ becomes
\begin{equation}
\mathcal{H}=\nu v_x
\begin{pmatrix}
\phi_{\text{eff}}(x) & -i\mathcal{L}^{-} \\
i\mathcal{L}^{+} & \phi_{\text{eff}}(x)
\end{pmatrix},
\label{eq:4}
\end{equation}
being
\begin{align}
\mathcal{L}^{\pm} &= \mp\frac{\mathrm{d}}{{\mathrm{d}}x} + W(x), \quad W(x) = \frac{v_y}{v_x}\left(k + \mathcal{A}_y(x)\right),
\nonumber \\
\phi_{\text{eff}}(x) &= \frac{v_t}{v_x} \left(k + \mathcal{A}_y(x) - \frac{\nu}{v_t}\phi(x)\right).
\label{eq:5}
\end{align}
In this way, the eigenvalue equation leads to the following coupled system of differential equations
\begin{equation}
\mathcal{L}^{\pm}\psi^{\pm}(x) = \nu\left(\frac{E}{v_x}- \nu\phi_{\text{eff}}(x)\right)\psi^{\mp}(x).
\label{eq:6}
\end{equation}
Moreover, the actions of the commutator and anticommutator between $\phi_{\text{eff}}(x),\mathcal{L}^{\pm}$ on the functions $\psi^{\pm}$ are given by:
\begin{align}
\left[ \phi_{\text{eff}}(x),\mathcal{L}^{\pm}\right]\psi^{\pm}(x)&= \phi^{'}_{\text{eff}}(x)\psi^{\pm}(x),\nonumber\\
\left\lbrace \phi_{\text{eff}}(x),\mathcal{L}^{\pm}\right\rbrace\psi^{\pm}(x)&=2\phi_{\text{eff}}(x)\left(\frac{E}{\nu v_x}-\phi_{\text{eff}}(x)\right)\psi^{\mp}(x)\mp\phi^{'}_{\text{eff}}(x)\psi^{\pm}(x).
\label{eq:7}
\end{align}
Now, with the aim of decoupling this system, we will consider the eigenvalue equation for the square of the Hamiltonian in Equation \eqref{eq:4}, which leads to 
\begin{align}
\mathcal{H}^2\bar{\Psi}(x)&=v_x^2
\begin{pmatrix}
H^++\phi^2_{\text{eff}}(x) & -i\left\lbrace \phi_{\text{eff}}(x),\mathcal{L}^{-}\right\rbrace
\\
i\left\lbrace \phi_{\text{eff}}(x),\mathcal{L}^{+}\right\rbrace & H^-+\phi^2_{\text{eff}}(x)
\end{pmatrix}
\begin{pmatrix}
\psi^+(x)\\
i\psi^-(x)
\end{pmatrix}\nonumber\\
&=E^2\bar{\Psi}(x)\nonumber\\
&=E^2\begin{pmatrix}
\psi^+(x)\\
i\psi^-(x)
\end{pmatrix},
\label{eq:8}
\end{align}
where $H^{\pm}$ are two one-dimensional Schrödinger Hamiltonian operators that factorize as:
\begin{equation}
H^{\pm} = \mathcal{L}^{\mp}\mathcal{L}^{\pm} = -\frac{{\rm d}^2}{{\rm d}x^2} + V^{\pm}(x),\quad \text{with}\quad
V^{\pm}(x) = W^2(x) \pm W'(x).
\label{eq:9}
\end{equation}
After some algebraic manipulations, Equation \eqref{eq:8} transforms into the following system of differential equations.
\begin{equation}
\left[H^{\pm}-\epsilon\right]\psi^{\pm}(x)=\pm\phi'_{\text{eff}}(x)\psi^{\mp}(x),
\quad\epsilon=\left(\frac{E}{v_x}-\nu\phi_{\text{eff}}(x)\right)^2,
\label{eq:10}
\end{equation}
Note that Equation \eqref{eq:10} leads us to a system of second-order differential equations that still remains coupled because functions $\phi'_{\text{eff}}(x)\psi^{\mp}(x)$ act as a source-like term. Nevertheless, to make the term that couples the differential system vanish, we can choose $\phi'_{\text{eff}}(x) = 0$, which can be guaranteed if the amplitudes of the electric and magnetic fields satisfy the relationship:
\begin{equation}
\mathcal{E}(x) = -\nu v_t \mathcal{B}(x).
\label{eq:11}
\end{equation}
In this scenario, the effective potential becomes constant:
\begin{equation}
\phi_{\text{eff}}(x) = \frac{v_t}{v_x} k.
\label{eq:12}
\end{equation}
Note that the suitable choice of the scalar potential $\phi(x)$ not only allows us to decouple the system of differential equations but also enables us to express information about the tilting of the cones as an energy level shift. This results in creating a bandgap between the conduction and valence bands.

In such a manner, the system of differential equations in \eqref{eq:10} decouples and becomes two eigenvalue equations for the Hamiltonians $H^{\pm}$ with eigenvalue $\epsilon$, as follows:
\begin{equation}
H^{\pm}\psi^{\pm}(x) = \epsilon\psi^{\pm}(x).
\label{eq:13}
\end{equation}
Furthermore, from Equation \eqref{eq:9}, it follows that the Hamiltonians $H^{\pm}$, and the operators $\mathcal{L}^{\pm}$ satisfy the following relationships.
\begin{equation}
H^{\pm}\mathcal{L}^{\mp} = \mathcal{L}^{\mp}H^{\mp}.
\label{eq:14}
\end{equation}
This relationship indicates that $H^{\pm}$ are supersymmetric partner Hamiltonians \cite{Kuru2009,Midya_2014,FernándezC.2022}. Consequently, the eigenvalues $\epsilon^+_n$ and eigenfunctions $\psi^+_n(x)$ of $H^+$ can be determined if those of $H^-$, i.e., $\epsilon^-_n$ and $\psi^-_n(x)$, are known, and vice versa. This can lead to one of the following three cases:
\begin{itemize}
\item [\bf{i)}] If $\mathcal{L}^{-}\psi^{-}_0(x)=0$, then $\epsilon_0^-=0$ and 
\begin{equation}
\psi^{+}_{n}(x) = \frac{\mathcal{L}^{-}\psi^{-}_{n+1}(x)}{\sqrt{\epsilon^{-}_{n+1}}},\quad\mbox{for}\quad
n=0,1,...
\label{eq:15}
\end{equation}
with eigenvalue $\epsilon_n^+=\epsilon_{n+1}^-$. Therefore, the eigenfunctions and eigenvalues of the Hamiltonian $\mathcal{H}$ are given by 
\begin{equation}
\Psi_n(x,y)=\frac{{\rm e}^{iky}}{\sqrt{2^{1-\delta_{n,0}}}}
\begin{pmatrix}
\left(1-\delta_{n,0}\right)\psi_{n-1}^+(x)\\
i\psi_n^-(x)
\end{pmatrix},\quad
E_n=\nu
v_{\rm t}k+\kappa v_x\sqrt{\epsilon_n^-}\quad\mbox{for}\quad
n=0,1...
\label{eq:16}
\end{equation}
being $\kappa=\pm1$ the band index for electrons ($1$) and holes ($-1$), respectively. 
\item [\bf{ii)}] If $\mathcal{L}^{+}\psi^{+}_0(x)=0$, then $\epsilon_0^+=0$ and 
\begin{equation}
\psi^{-}_{n}(x) = \frac{\mathcal{L}^{+}\psi^{+}_{n+1}(x)}{\sqrt{\epsilon^{+}_{n+1}}},\quad\mbox{for}\quad
n=0,1,...
\label{eq:17}
\end{equation}
with eigenvalue $\epsilon_n^-=\epsilon_{n+1}^+$. Therefore, the eigenfunctions and eigenvalues of the Hamiltonian $\mathcal{H}$ are given by 
\begin{equation}
\Psi_n(x,y)=\frac{{\rm e}^{iky}}{\sqrt{2^{1-\delta_{n,0}}}}
\begin{pmatrix}
\psi_n^+(x)\\
i\left(1-\delta_{n,0}\right)\psi_{n-1}^-(x)
\end{pmatrix},\quad
E_n=\nu
v_{\rm t}k+\kappa v_x\sqrt{\epsilon_n^+}\quad\mbox{for}\quad
n=0,1...
\label{eq:18}
\end{equation}
\item [\bf{iii)}] If $\mathcal{L}^{+}\psi^{+}_0(x)\neq0$ and $\mathcal{L}^{-}\psi^{-}_0(x)\neq0$, then the ground state of both Hamiltonians $H^{\pm}$ has energy different from zero. Moreover, it is fulfilled that 
\begin{equation}
\psi^{\pm}_{n}(x) = \frac{\mathcal{L}^{\mp}\psi^{\mp}_{n}(x)}{\sqrt{\epsilon_{n}}},\quad\mbox{for}\quad
n=0,1,...
\label{eq:19}
\end{equation}
with eigenvalue $\epsilon_n^{-}=\epsilon_{n}^+=\epsilon_n$. Therefore, the eigenfunctions and eigenvalues of the Hamiltonian $\mathcal{H}$ are given by 
\begin{equation}
\Psi_n(x,y)=\frac{{\rm e}^{iky}}{\sqrt{2}}
\begin{pmatrix}
\psi_n^+(x)\\
i\psi_{n}^-(x)
\end{pmatrix},\quad
E_n=\nu
v_{\rm t}k+\kappa v_x\sqrt{\epsilon_n}\quad\mbox{for}\quad
n=0,1...
\label{eq:20}
\end{equation}
\end{itemize}
We must emphasize that each of these cases depends solely on the choice of the vector potential that determines the function $W$ and, therefore, on the operators $\mathcal{L}^{\pm}$.

From now on and unless otherwise indicated, in this work, we will focus on constant field profiles that lead to eigenfunctions and eigenvalues similar to those obtained in the previous cases.
\subsection{Constant magnetic and electric fields}\label{1.1}
Let us consider a material with an effective Hamiltonian $\mathcal{H}$ as in Equation \eqref{eq:2} and localized in the presence of constant magnetic and electric fields given by:
\begin{equation}
\vec{\mathcal{B}}=\mathcal{B}_0\hat{e}_z,\quad
\vec{\mathcal{E}}=-\nu v_{t}\mathcal{B}_0\hat{e}_x,
\label{eq:23}
\end{equation}
where $\mathcal{B}_0$ is a positive constant, then the scalar an vector potential are given by 
\begin{equation}
\vec{\mathcal{A}}=\mathcal{B}_0x\hat{e}_y,\quad
\phi(x)=\nu v_{t}\mathcal{B}_0x.
\label{eq:24}
\end{equation}
This ensures that Equation \eqref{eq:9} is satisfied, and the components $\psi_n^{\pm}(x)$ of the eigenstates $\Psi_n$ are identified as eigenfunctions of the supersymmetric Hamiltonians $H^{\pm}$, which turn out to be 
\begin{equation}
H^{\pm}=-\frac{\mbox{d}^2}{\mbox{d}x^2}
+\frac{\omega^2}{4}\left(x+\frac{2\tilde{k}}{\omega}\right)^2\pm
\frac{\omega}{2},
\label{eq:25}
\end{equation}
where $\omega$ and $\tilde{k}$ have been defined as
\begin{equation}
\omega=2\frac{v_y}{v_x}\mathcal{B}_0,\quad\tilde{k}=\frac{v_y}{v_x}k.
\label{eq:26}
\end{equation}
In this way the components $\psi_n^{\pm}$ and the eigenvalue $\epsilon_n^-$ turn out to be 
\begin{equation}
\psi^{\pm}_n(x) = \sqrt{\frac{1}{2^n n!}\left(\frac{\omega}{2\pi}\right)^{\frac{1}{2}}}\,\mathrm{H}_n\left[\sqrt{\frac{\omega}{2}}\left(x + \frac{2\tilde{k}}{\omega}\right)\right]{\rm e}^{-\frac{\omega}{4}\left(x + \frac{2\tilde{k}}{\omega}\right)^{2}},\quad
\epsilon_n^- = n\omega,\quad
n = 0,1,...
\label{eq:27}
\end{equation}
where $\mathrm{H}_n(z)$ represents the Hermite polynomial of degree $n$. As a result, the eigenfunctions of the Hamiltonian $\mathcal{H}$ are given by
\begin{equation}
\Psi_n(x,y) = \frac{{\rm e}^{iky}}{\sqrt{2^{1-\delta_{n0}}}}
\begin{pmatrix} 
(1 - \delta_{n0}) \psi^{-}_{n-1}(x)\\ 
\\
i \psi_n^-(x)
\end{pmatrix},\quad\text{for}\quad
n = 0,1,...
\label{eq:28}
\end{equation}
and the corresponding eigenvalues are $E_n =\nu v_tk+\kappa v_x\sqrt{n\omega}$. Note that since $\psi_n^{-}$ are the eigenfunctions of the shifted harmonic oscillator, then the set of well-known one-dimensional ladder operators $\left\lbrace\Theta^{-},\Theta^{+},N\right\rbrace$ is given by:
\begin{align}
\Theta^{-}=&\frac{1}{\sqrt{2}}\left(\zeta + \frac{\rm d}{{\rm d}\zeta}\right),\nonumber\\
\Theta^{+}=&\frac{1}{\sqrt{2}}\left(\zeta - \frac{\rm d}{{\rm d}\zeta}\right),\nonumber\\
N=&\Theta^{+}\Theta^{-},
\label{eq:29}
\end{align}
where $\zeta = \sqrt{\frac{\omega}{2}}\left(x + \frac{2\tilde{k}}{\omega}\right)$, such that 
\begin{align}
\Theta^{-}\psi_n^{-}(x)&=\sqrt{n}\psi_{n-1}^{-}(x),\nonumber\\
\Theta^{+}\psi_n^{-}(x)&=\sqrt{n+1}\psi_{n+1}^{-}(x),\nonumber\\
N\psi_n^{-}(x)&=n\psi_{n}^{-}(x).
\label{eq:30}   
\end{align}
\section{Coherent sates}\label{2}
\begin{figure}[h!] 
\begin{center}
\includegraphics[width=0.6\textwidth]{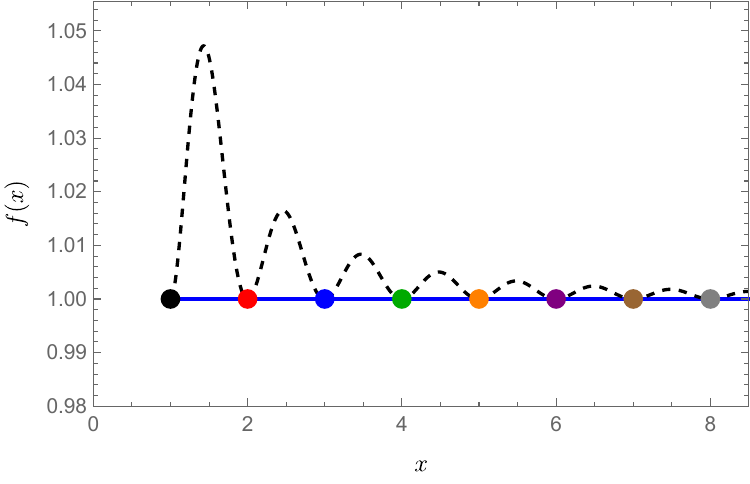}
\caption{Plot of two different choices of the function $f$. The solid line represents $f(x)=1$, while the dashed line represents $f(x)=\left(\frac{\sin{(\pi x)}}{\pi x}\right)^2+1$. It can be seen how $f_n$ is the same for both expressions since the values for $n\in\mathbb{Z}^+$ always have a value of 1.}
\label{Figure1}
\end{center}
\end{figure}

To build now the coherent states associated with this system, we need to introduce an appropriate $2\times2$ annihilation operator $A^-$ for those eigenfunctions in Equation \eqref{eq:28}. For that reason we will define $A^-$ as follows \cite{FernándezC.2022}:
\begin{equation}
A^{-}=
\begin{pmatrix} 
\mathcal{L}^{-}\frac{1}{\sqrt{N}}\Theta^{-}\sqrt{\frac{f(N)}{N}}
\mathcal{L}^{+} & -i\mathcal{L}^{-}\frac{1}{\sqrt{N}}\Theta^{-}\sqrt{f(N)}\\ 
&\\
i\Theta^{-}\sqrt{\frac{f(N)}{N}}\mathcal{L}^{+} & \Theta^{-}\sqrt{f(N)}
\end{pmatrix},
\label{eq:31}
\end{equation}
where \(f\) is any real and positive function with a well-defined Taylor series. Then, the action of \(A^{-}\) on \(\Psi_n\) is as follows:
\begin{equation}
A^{-}\Psi_n(x,y)=2\sqrt{nf_n}\Psi_{n-1}(x,y)
\begin{cases}
0\quad\text{for}\quad n=0,\\
\\
\frac{1}{\sqrt{2}}\quad\text{for}\quad n=1,\\
\\
1\quad\text{for}\quad n\geq2.
\end{cases}
\label{eq:32}
\end{equation}
where $f_n$ is given by the values of the function $f$ when the argument is a positive integer, i.e., $f(n):=f_n$. Note that this proposal actually represents a family of annihilation operators labeled by $f$. However, we must emphasize that $f$ takes a relevant role when its argument is a positive integer. For this reason, two different functions of a real variable will represent the same annihilation operator if the same values are obtained for all $n\in\mathbb{Z}^+$, as shown in Figure \ref{Figure1}.
\subsection{The Barut-Girardello coherent states}\label{2.1}
There are different definitions that allow us to generalize the concept of coherent states (CS) of the harmonic oscillator. Nevertheless, in this work, we consider a coherent state as an eigenstate of the annihilation operator with complex eigenvalue $\alpha$. This type of coherent state is also known as the Barut-Girardello coherent state (BGCS).\\
\\
Let $\Psi_{\alpha}(x,y)$ be the Barut-Girardello coherent state, such that
\begin{equation}
A^{-}\Psi_{\alpha}(x,y)=\alpha\Psi_{\alpha}(x,y),
\label{eq:33}
\end{equation}
where \(\alpha\) is the complex eigenvalue. Since $\Psi_{\alpha}$ can be represented in the basis of Hamiltonian eigenfunctions 
\begin{equation}
\Psi_{\alpha}(x,y)=\sum_{n=0}^{\infty}d_n\Psi_n(x,y).
\label{eq:34}
\end{equation}
By substituting \eqref{eq:34} into \eqref{eq:33} and performing some algebraic manipulations, we get the following recurrence relation
\begin{equation}
\alpha d_{n}=\frac{2\sqrt{(n+1)f_{n+1}}d_{n+1}}{\sqrt{2^{\delta_{n,0}}}}, \quad \text{for}\quad n=0,1,...
\label{eq:35}
\end{equation}
which implies that
\begin{equation}
d_n=\frac{\alpha^{n}d_0}{2^n\sqrt{n!\left[f_n\right]!}}\sqrt{2^{1-\delta_{n,0}}}.
\label{eq:36}
\end{equation}
Here, the generalized factorial of any arbitrary function $f$ with integer argument $n$ (remember that $f_n:=f(n)$) has been defined as
\begin{equation}
\left[f_n\right]!=
\begin{cases}
1\quad\text{for}\quad n=0,\\
\\
f_1f_2\cdot\cdot\cdot f_{n-1} f_n\quad\text{for}\quad n\geq1.
\end{cases}
\label{eq:37}
\end{equation}
Thus, the normalized Barut-Girardello coherent state turns out to be:
\begin{equation}
\Psi_{\alpha}(x,y)=\mathcal{C}_{\alpha}\sum_{n=0}^{\infty}
\frac{\alpha^{n}}{2^n\sqrt{n!\left[f_n\right]!}}\sqrt{2^{1-\delta_{n,0}}}\Psi_n(x,y),
\label{eq:38}
\end{equation}
being $\mathcal{C}_{\alpha}$ a normalization constant given by
\begin{equation}
\mathcal{C}_{\alpha}=
\left(\sum_{n=0}^{\infty}
\frac{|\alpha|^{2n}}{2^{2n}n!\left[f_n\right]!}2^{1-\delta_{n,0}}\right)^{-\frac{1}{2}}.
\label{eq:39}
\end{equation}
It is important to highlight certain key aspects regarding the Barut-Girardello coherent states derived above:
\begin{itemize}
\item [•] While $f_n$ was initially regarded as an arbitrary real and positive function, it is crucial for it to ensure the convergence of the state in Equation \eqref{eq:38}. Consequently, this constrains the family of 'well-behaved' annihilation operators.
\item [•] Although the definition of the Barut-Girardello coherent state is not restricted to systems with an infinite-dimensional basis, for those with finite dimensions, the coherent states in \eqref{eq:38} will coincide with the ground state, and the corresponding eigenvalue will be zero.
\end{itemize}
Next, we will some examples using the fidelity and the Wigner function to examine the temporal and phase-space behaviors of the Barut-Girardello coherent states of the Equation \eqref{eq:38}.
\section{Fidelity and Wigner function}
\label{3}
The coherent states obtained in Equation \eqref{eq:38} only describe the initial state of a particle at time $t=0$. In order to obtain the BGCS at an arbitrary time $t$ it is necessary to apply the temporal evolution operator $\mathcal{U}(t)=\exp(-i\mathcal{H}t)$ to these initial coherent states. Then, the BGCS for any arbitrary time is given by  
\begin{equation}
\Psi_{\alpha}(x,y,t)=\mathcal{C}_{\alpha}
{\rm e}^{-i\nu v_tk t}
\sum_{n=0}^{\infty}
\frac{\alpha^{n}{\rm e}^{-i\kappa v_x\sqrt{n\omega} t}}{2^n\sqrt{n!\left[f_n\right]!}}\sqrt{2^{1-\delta_{n,0}}}\Psi_n(x,y).
\label{eq:40}
\end{equation}
Although the valley index $\nu$ plays an important role in some of the physical properties of the system such as the probability current, Equation \eqref{eq:40} shows us that in the case of temporal evolution this is not the case since factor ${\rm e}^{-i\nu v_tk t}$ is only a global phase.
\subsection{Fidelity}\label{3.1}
A method that allows us to determine the possible existence of evolution periods is through fidelity, which is defined as:
\begin{equation}
F(\phi,\xi) \equiv |\braket{\phi|\xi}|^2.
\label{eq:41}
\end{equation}
Thus, fidelity allows us to know the similarity between two states, having the greatest similarity when $F(\phi,\xi) = 1$. In that case, it can be said that $\ket{\phi}, \ket{\xi}$ differ at most by a global phase factor and both represent the same quantum state. Furthermore, if the fidelity between an initial state and its evolved state is calculated, the result will allow us to calculate the values of $t$ for which it is closest to the initial state.

In this way, after calculating the fidelity with the initial BGCS in Equation \eqref{eq:38} and its evolved counterpart in \eqref{eq:40}, it is determined that
\begin{equation}
F(\Psi_{\alpha},\Psi_{\alpha}(t)) = 
\mathcal{C}^4_{\alpha} 
\sum_{n,m=0}^{\infty} \frac{|\alpha|^{2(n+m)}}{2^{2(n+m)}n!m!\left[f_n\right]!\left[f_m\right]!} 2^{2-\delta_{n,0}-\delta_{m,0}}\cos\left[\left (\sqrt{n}-\sqrt{m}\right)\sqrt{\omega}v_xt\right].
\label{eq:42}
\end{equation}
The preceding expression indicates that the fidelity between these two states does not depend on the phase of the complex eigenvalue $\alpha$ nor the band index $\kappa$ or the valley index $\nu$. These characteristics are illustrated in subsection \ref{3.3}.
\subsection{Wigner function}\label{3.2}
In quantum mechanics, the so-called Weyl quantization allows us to obtain an operator $\mathcal{O}$, associated with a classical observable $O$ defined as a function on phase space, as \cite{weyl1931theory}
\begin{equation}
    \mathcal{O}=\frac{1}{(2\pi)^{2n}}\int_{-\infty}^{\infty} O(\mathbf{r},\mathbf{p})\, {\rm e}^{i[\mathbf{z}\cdot(\mathbf{r}-\mathcal{R})+\mathbf{y}\cdot(\mathbf{p}-\mathcal{P})]}\,{\rm d}\mathbf{z}\,{\rm d}\mathbf{y}\,{\rm d}\mathbf{r}\,{\rm d}\mathbf{p},
\label{eq:43}
\end{equation}
where \(\mathbf{r} = (r_1, r_2, ..., r_n)\) and \(\mathbf{p} = (p_1, p_2, ..., p_n)\) are \(n\)-dimensional vectors representing the classical phase-space position and momentum values, $\mathbf{y}$ and $\mathbf{z}$ are $n$-dimensional vectors with the dimension of position and momentum, respectively, as well $\mathcal{R}$ and $\mathcal{P}$ are position and momentum operators, respectively.

Inversely, the Wigner map $O$ of the operator $\mathcal{O}$ in a given Hilbert space $\mathcal{H}$ is defined by \cite{Case2008}
\begin{equation}
O(\mathbf{r}, \mathbf{p}) = \frac{1}{(2\pi)^n} \int_{-\infty}^{\infty}
{\rm e}^{i\mathbf{p}\cdot\mathbf{r'}} \langle \mathbf{r}-\frac{\mathbf{r'}}{2}| \mathcal{O} | \mathbf{r}+\frac{\mathbf{r'}}{2} \rangle \, {\rm d}\mathbf{r'},
\label{eq:44}
\end{equation}
This Weyl transform converts an operator $\mathcal{O}$ into a function $O$ of $\mathbf{r}$ and $\mathbf{p}$.

Therefore, to analyze the distribution of the coherent states in phase space, we can utilize the time-dependent Wigner function $W(\mathbf{r}, \mathbf{p},t)$, which defines a quasiprobability distribution as follows:
\begin{equation}
W(\mathbf{r}, \mathbf{p},t) = \frac{1}{(2\pi)^n} \int_{-\infty}^{\infty}
{\rm e}^{i\mathbf{p}\cdot\mathbf{r'}} \langle \mathbf{r}-\frac{\mathbf{r'}}{2}| \rho | \mathbf{r}+\frac{\mathbf{r'}}{2} \rangle \, {\rm d}\mathbf{r'},
\label{eq:45}
\end{equation}
where $\rho=\ket{\Psi_{\alpha}(t)}\bra{\Psi_{\alpha}(t)}$ is the density matrix. The above expression can be interpreted as the Fourier transform of the density matrix. Furthermore, defining the quantities
\begin{equation}
    \mathbf{r}_i=\mathbf{r}-\frac{\mathbf{r'}}{2}, \quad \mathbf{r}_f=\mathbf{r}+\frac{\mathbf{r'}}{2},
\label{eq:46}
\end{equation}
the parameter $\mathbf{r}=\frac{\mathbf{r}_i+\mathbf{r}_f}{2}$ is the mean position of a particle in the interval of time $\Delta t=t_f-t_i$, while $\mathbf{r'}=\mathbf{r}_f-\mathbf{r}_i$ measures the displacement of the particle in such a interval of time, and $\mathbf{p}$ identifies the momentum along the period $\Delta t$. Then, by considering that the wave function for the BGCS in Equation \eqref{eq:40} is defined as $\Psi_{\alpha}(x,y,t)=\braket{\mathbf{r}|\Psi_{\alpha}(t)}$, the resulting Wigner function can be written as
\begin{align}
W_{\alpha}(\mathbf{r}, \mathbf{p},t)&=W_{\alpha}(y, p_y)\times
W_{\alpha}(x,p_x,t)\nonumber\\
&=\delta(p_y-k)\times \frac{\mathcal{C}_{\alpha}^2}{2\pi}\sum_{n,m=0}^{\infty} \frac{|\alpha|^{n+m} \exp\left[i(n-m)\theta+i(\sqrt{m}-\sqrt{n})\kappa v_x\sqrt{\omega}t\right]}{2^{n+m}\sqrt{n!m!\left[f_n\right]!\left[f_m\right]!}} \mathcal{M}_{n,m}(x,p_x),
\label{eq:47}
\end{align}
where $\theta$ is a real number such as $\alpha=|\alpha|{\rm e}^{i\theta}$, $\mathcal{M}_{n,m}(x,p_x)$ is a $2\times2$ matrix given by
\begin{equation}
\mathcal{M}_{n,m}(x,p_x)=
\begin{pmatrix} 
(1-\delta_{n0})(1-\delta_{m0}) I_{n-1,m-1}(x,p_x) & -i(1-\delta_{n0})I_{n-1,m}(x,p_x)\\ 
&\\
i(1-\delta_{m0})I_{n,m-1}(x,p_x) & I_{n,m}(x,p_x)
\end{pmatrix},
\label{eq:48}
\end{equation}
and
\begin{equation}
I_{n,m}(x,p_x)=
2\int_{-\infty}^{\infty}
\mbox{e}^{2ip_x z}
\psi_n^{\pm}(x-z)
\psi_m^{\pm}(x+z)\mbox{d}z,
\label{eq:49}
\end{equation}
by consider the functions $\psi_n^\pm(x)$ from Equation \eqref{eq:27}, $I_{n,m}(x,p_x)$ reduces to
\begin{equation}
I_{n,m}(x,p_x)=2{\rm e}^{-\frac{1}{2}|u|^2}\times\left\lbrace
\begin{array}{c}
(-1)^m \bar{u}^{n-m}\sqrt{\frac{m!}{n!}} \mbox{L}_m^{n-m}(|u|^2)\quad\mbox{for}\quad
n\geq m,\\
\\
(-1)^n u^{m-n}\sqrt{\frac{n!}{m!}} \mbox{L}_n^{m-n}(|u|^2)\quad\mbox{for}\quad
n<m,
\end{array}\right.
\label{eq:50}
\end{equation}
being $\mbox{L}_n^{m-n}$ the generalized Laguerre polynomials and $u=\sqrt{\omega}\left[\left(x+\frac{2\tilde{k}}{\omega}\right)+\frac{2ip}{\omega}\right]$. The Wigner function in Equation \eqref{eq:47} does not depend explicitly on the valley index $\nu$, but it does depend on the band index $\kappa$ in contrast to the fidelity of Equation \eqref{eq:42}. On the other hand, we can observe that the coherent state $\Psi_{\alpha}$ possesses a spinorial nature; this is why a matrix Wigner function is obtained. However, the quasiprobability distribution will be determined by the trace of the matrix from Equation \eqref{eq:47} which is an invariant. In the next part, we will choose some specific forms for $f_n$ and show the behavior of the fidelity and trace of the Wigner function under those choices. This will allow us to compare this behavior with that of the coherent states of the harmonic oscillator.
\subsection{Some examples}\label{3.3}
If we want to analyze the behavior of BGCS in Equations \eqref{eq:38} and \eqref{eq:40}, it is necessary to set a value for $f_n$, since its temporal evolution, and therefore the fidelity and Wigner function, depends strongly on this choice. Below we will show a couple of examples that allow us to manipulate the temporal and phase space behavior of the BGCS.
\subsubsection{Case \boldmath{$f_n=2^{\delta_{n,1}-2}p^2$}}\label{3.3.1}
\begin{figure}[h!] 
\begin{center}
\subfigure[]{\includegraphics[width=0.45\textwidth]{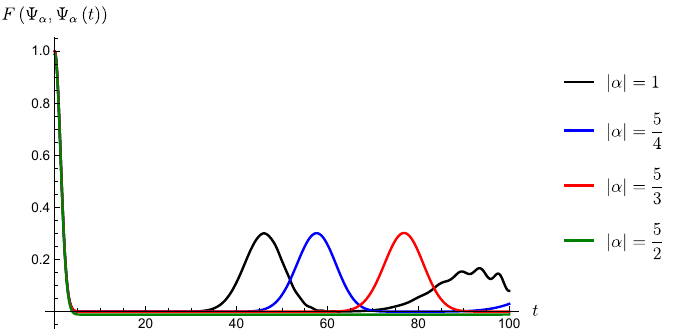}}
\subfigure[]{\includegraphics[width=0.45\textwidth]{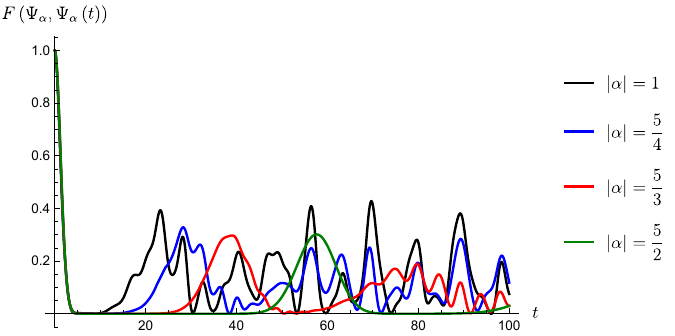}}
\subfigure[]{\includegraphics[width=0.45\textwidth]{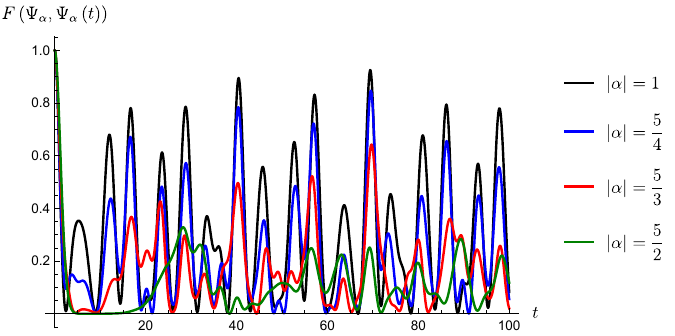}}
\subfigure[]{\includegraphics[width=0.45\textwidth]{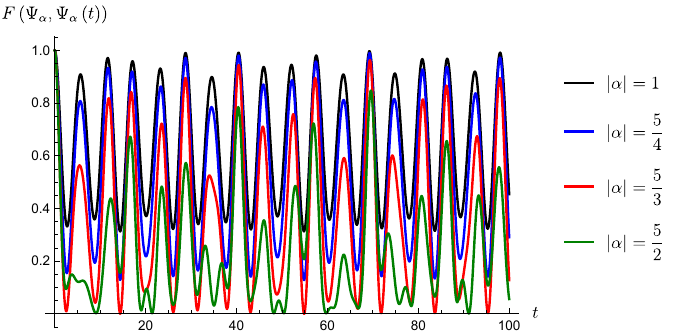}}
\caption{Fidelity plots for the BGCS of Equation \eqref{eq:52} and their time evolution for different values of the norm of the complex number $\alpha$ and (a) $p=\frac{1}{4}$, (b) $p=\frac{1}{2}$, (c) $p=1$ and (d) $p=2$. The parameters have been taken as $v_x=0.86$, $v_y=0.69$, $v_t=0.32$, $\mathcal{B}_0=1$ and $k=0$.}
\label{Figure2}
\end{center}
\end{figure}

\begin{figure}[h!] 
\begin{center}
\subfigure[]{\includegraphics[width=0.35\textwidth]{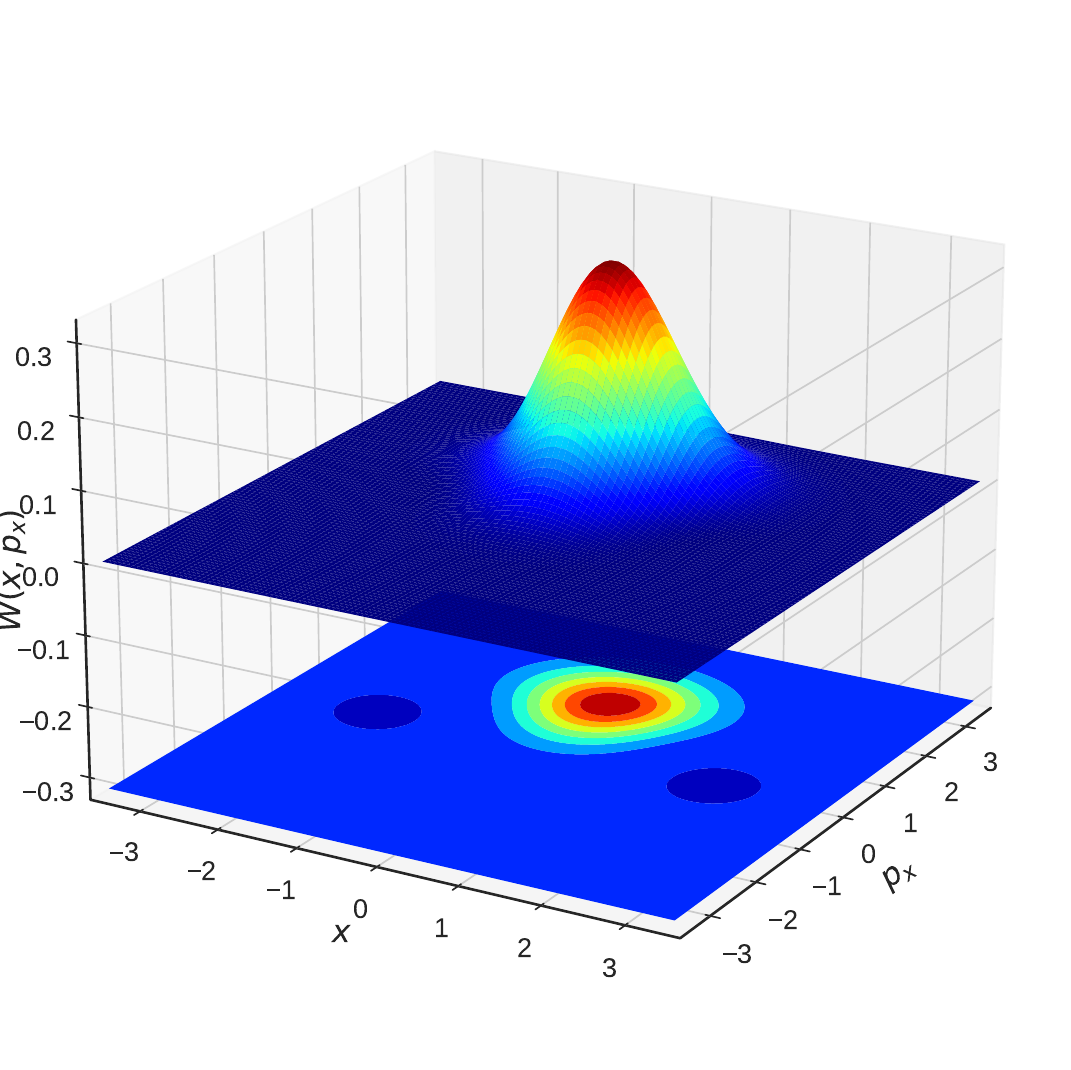}}\hspace{2cm}
\subfigure[]{\includegraphics[width=0.35\textwidth]{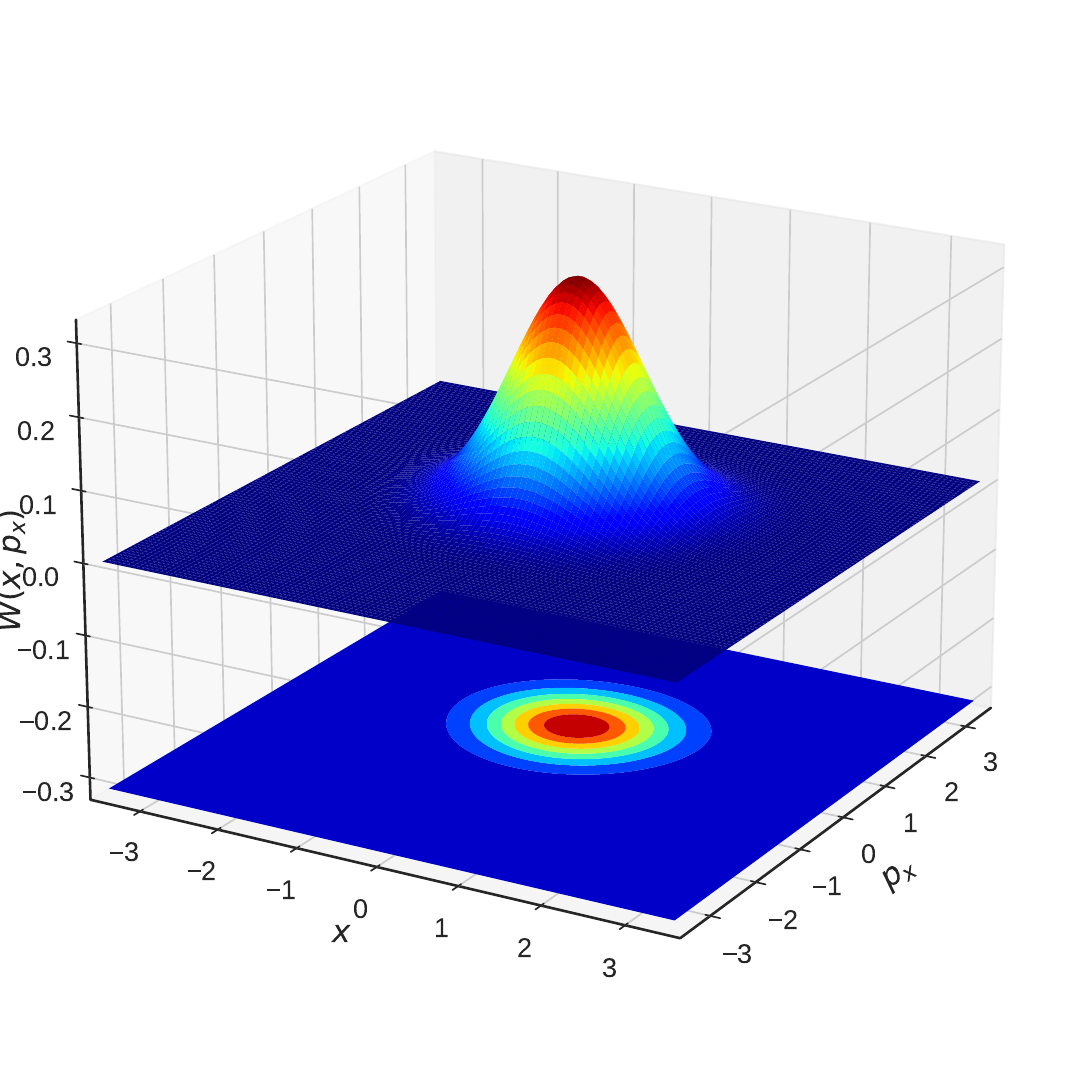}}
\subfigure[]{\includegraphics[width=0.35\textwidth]{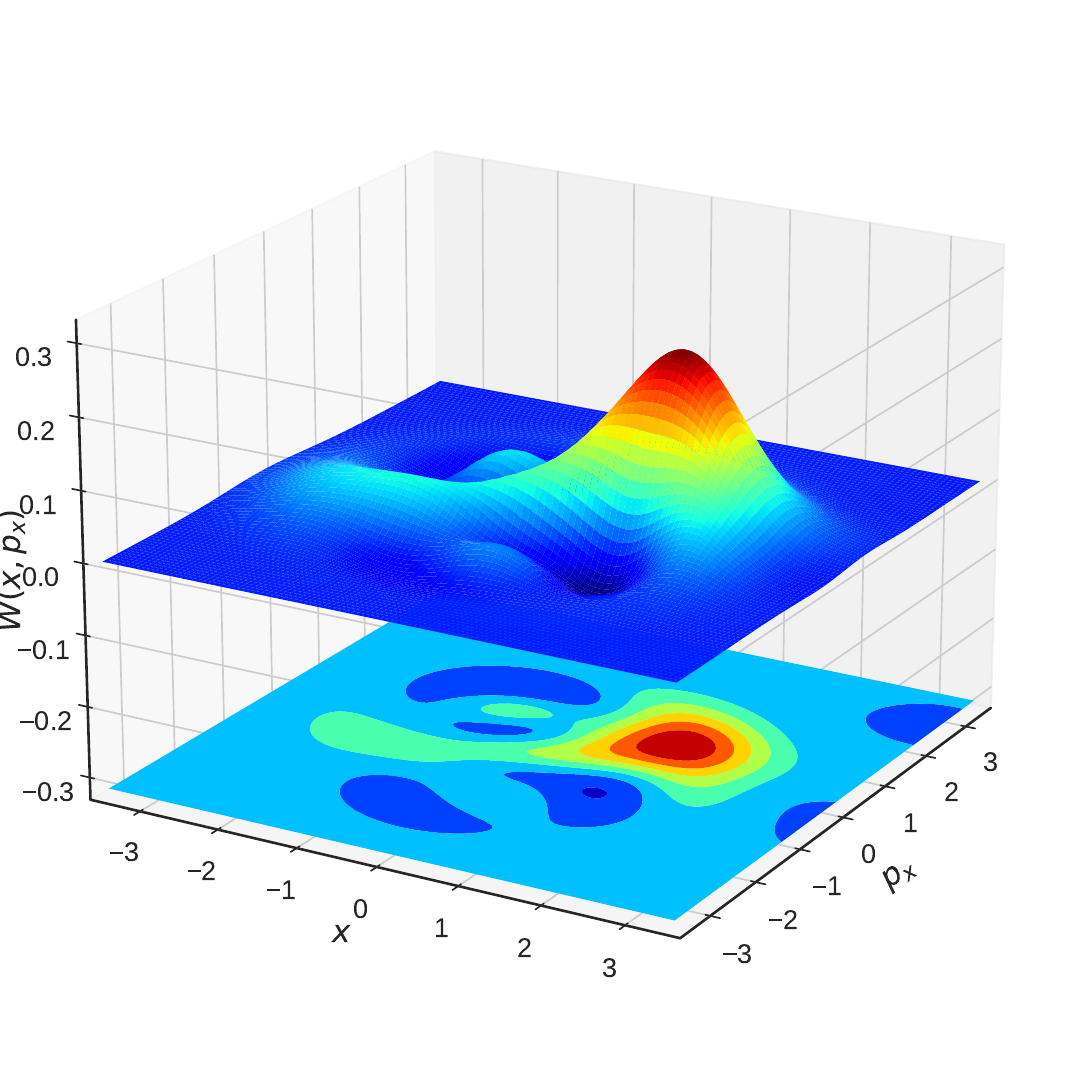}}\hspace{2cm}
\subfigure[]{\includegraphics[width=0.35\textwidth]{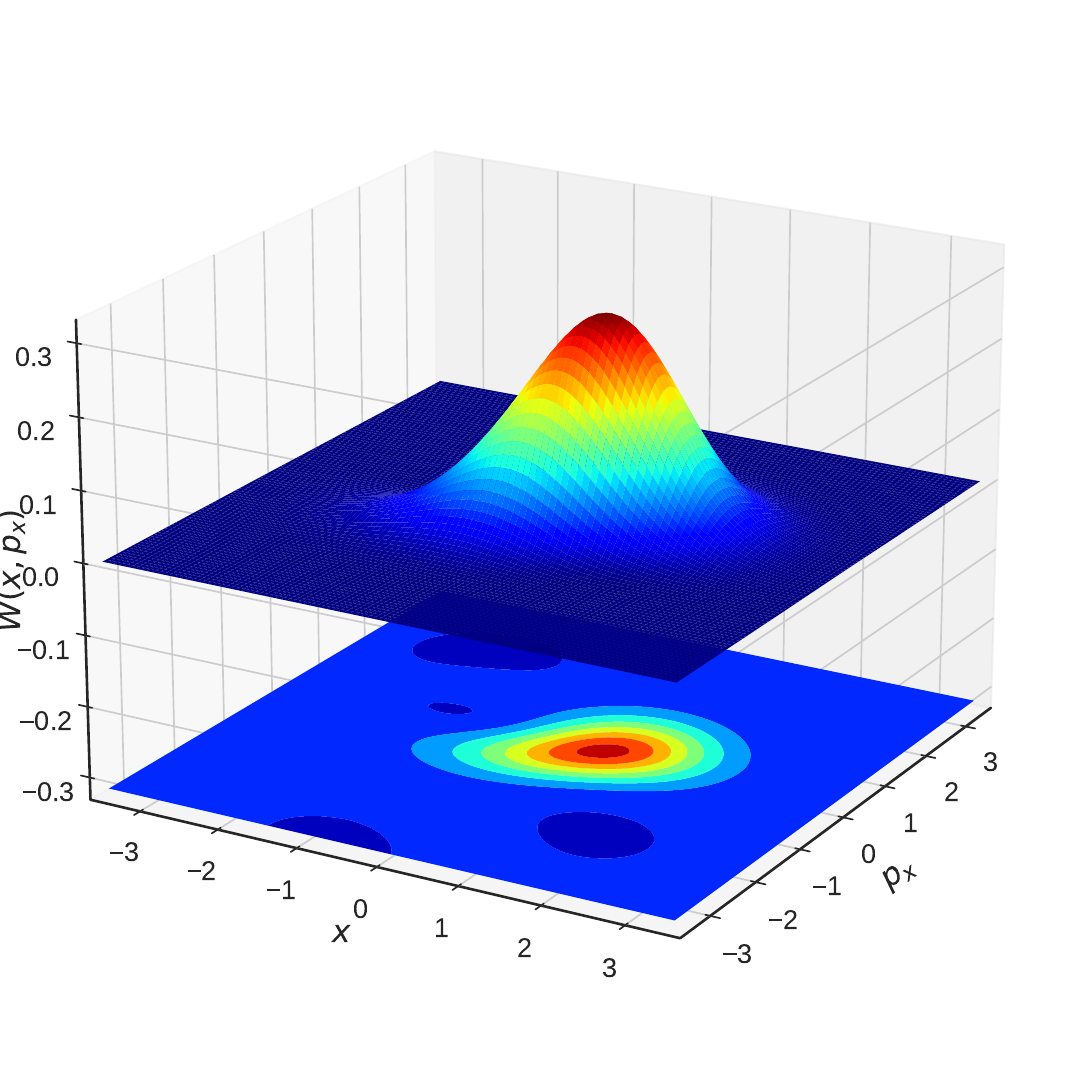}}
\subfigure[]{\includegraphics[width=0.35\textwidth]{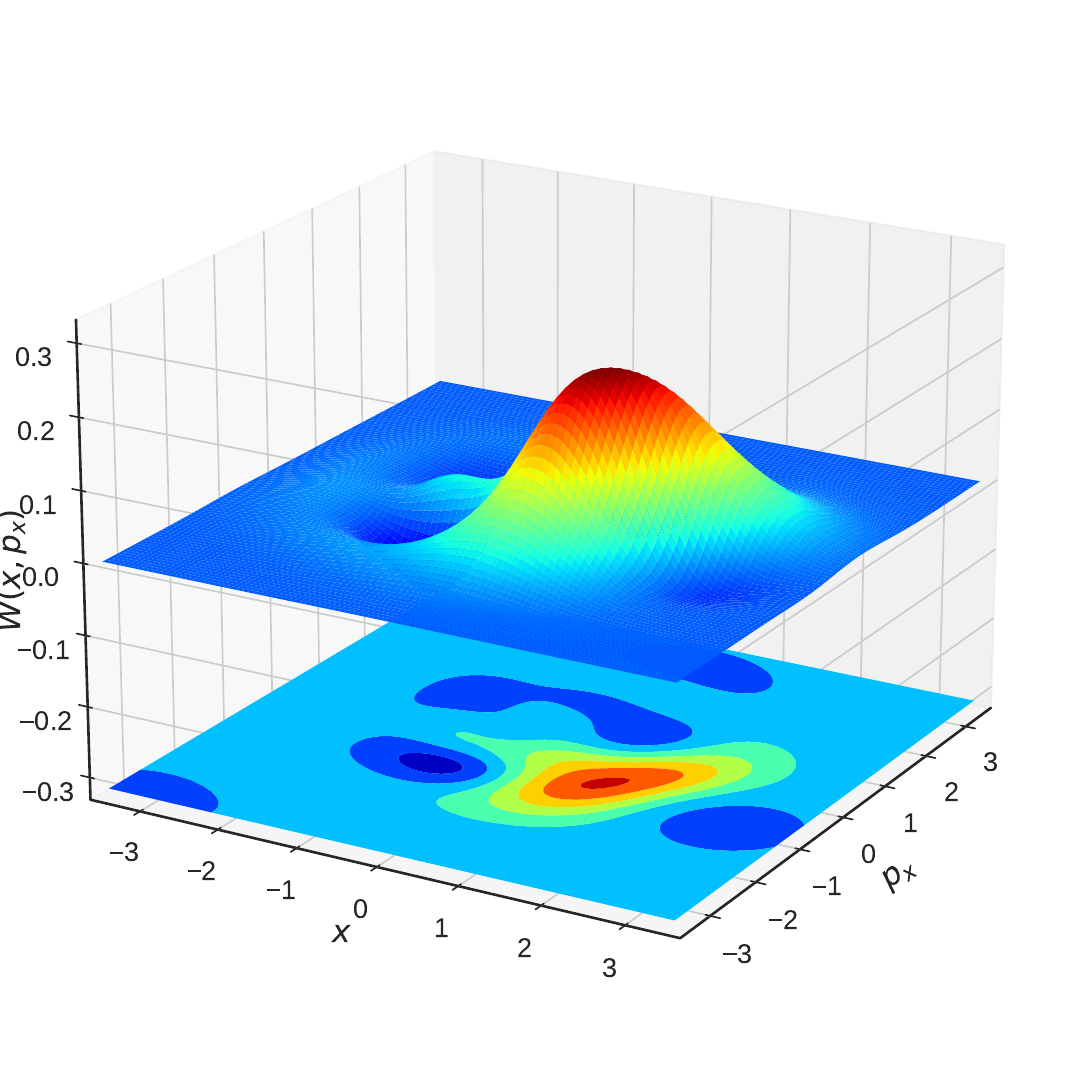}}\hspace{2cm}
\subfigure[]{\includegraphics[width=0.35\textwidth]{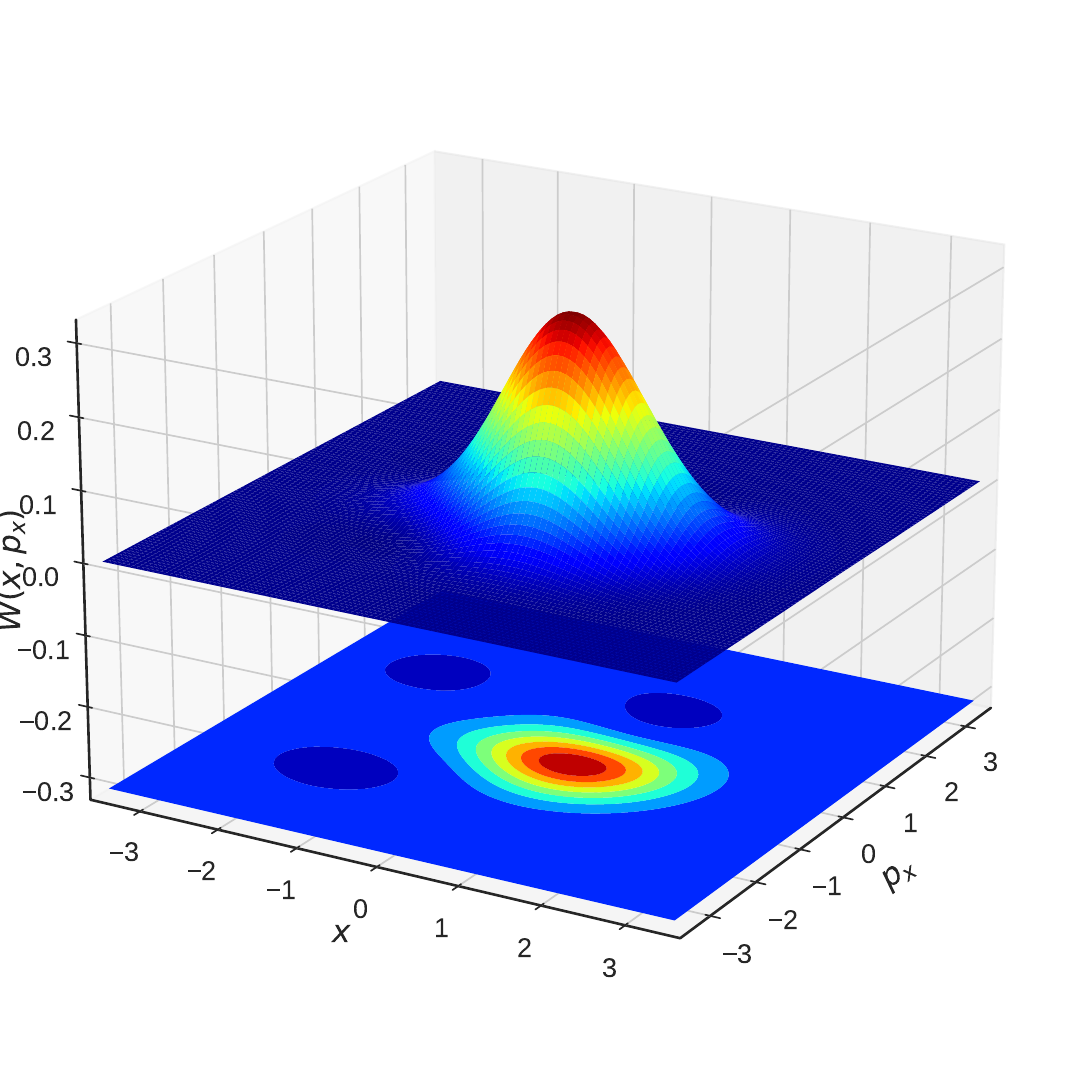}}
\caption{Plots of the trace of the Wigner function given in Equation \eqref{eq:54} with $\alpha=\frac{5}{4}{\rm e}^{i\pi/2}$ and $\kappa=+1$. The parameter $p$ has been taken as (left column) $p=1$ and (right column) $p=2$ for (a, b) $t=0$, (c, d) $t=42$, (e, f) $t=60$. The parameters have been taken as $v_x=0.86$, $v_y=0.69$, $v_t=0.32$, $\mathcal{B}_0=1$ and $k=0$.}
\label{Figure3}
\end{center}
\end{figure}

Let $f_n=2^{\delta_{n,1}-2}p^2$, with $p$ any positive real number. Then we can affirm that
\begin{equation}
\left[f_n\right]!=\frac{2^{1-\delta_{n,0}}p^{2n}}{2^{2n}},\quad
n=0,1,...
\label{eq:51}
\end{equation}
Therefore, the radius of convergence in the whole complex plane $\mathbf{C}$ is given by $|\alpha|=\infty$ and the normalization constant $\mathcal{C}_{\alpha}$ simplifies to $\mathcal{C}_{\alpha}={\rm e}^{-\frac{|\alpha|^2}{2p^2}}$. In this way, the Barut-Girardello coherent state of Equation \eqref{eq:38} can be written as:
\begin{equation}
\Psi_{\alpha}(x,y)={\rm e}^{-\frac{|\alpha|^2}{2p^2}}\sum_{n=0}^{\infty}
\frac{\alpha^{n}}{p^n\sqrt{n!}}\Psi_n(x,y).
\label{eq:52}
\end{equation}
Consequently, the fidelity between this state and the one corresponding to its time evolution is given by:
\begin{equation}
F(\Psi_{\alpha},\Psi_{\alpha}(t))={\rm e}^{-2\left(\frac{|\alpha|}{p}\right)^2}
\sum_{n,m=0}^{\infty}\left(\frac{|\alpha|}{p}\right)^{2(n+m)}\frac{\cos
\left[\left(\sqrt{n}-\sqrt{m}\right)\sqrt{\omega}v_x t\right]}{n!m!},
\label{eq:53}
\end{equation}
and the corresponding trace of 
the matrix Wigner function is
\begin{equation}
W_{\alpha}(\mathbf{r}, \mathbf{p},t)=\delta(p_y-k)\times \frac{{\rm e}^{-\left(\frac{|\alpha|}{p}\right)^2}}{2\pi}\sum_{n,m=0}^{\infty} 
\left(\frac{|\alpha|}{p}\right)^{n+m}
\frac{ {\rm exp}\left[(in-m)\theta+i(\sqrt{m}-\sqrt{n})\kappa v_x\sqrt{\omega}t\right]}{\sqrt{2^{2-\delta_{n0}-\delta_{m0}}n!m!}}{\rm Tr}\left[\mathcal{M}_{n,m}(x,p_x)\right].
\label{eq:54}
\end{equation}
From the previous expressions, the following can be stated: 1) The BGCS of Equation \eqref{eq:52} maintains the same algebraic form as the standard coherent states of the harmonic oscillator; thus, the Fock states $\Psi_n(x,y)$ maintain a Poisson distribution $P(n)=\vert a_{n}\vert^2={\rm e}^{-\left(\frac{|\alpha|}{p}\right)^{2}}\frac{|\alpha|^{2n}}{p^{2n}n!}$, with the mean equal to $|\alpha|^{2}$ and 2) the phase $\theta$ of the complex number $\alpha$ does not modify the value of the fidelity, but it does affect the Wigner function, which is also influenced by the band index $\kappa$. Figures \ref{Figure2} and \ref{Figure3} show the plots of these functions, as well as the effect that the parameter $p$ has on their behavior, allowing us to manipulate its value for a fixed time and complex label $\alpha$.
\subsubsection{Case \boldmath{$f_n=2^{\delta_{n,1}-2}\frac{p^2}{n}$}}\label{3.3.2}
\begin{figure}[h!] 
\begin{center}
\subfigure[]{\includegraphics[width=0.45\textwidth]{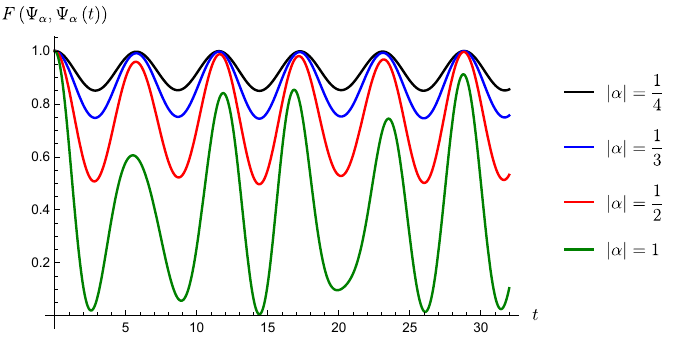}}
\subfigure[]{\includegraphics[width=0.45\textwidth]{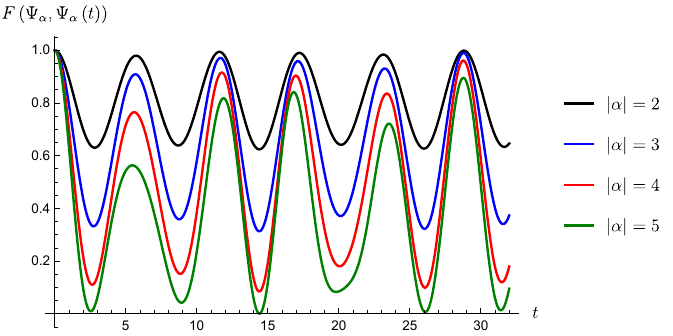}}
\subfigure[]{\includegraphics[width=0.45\textwidth]{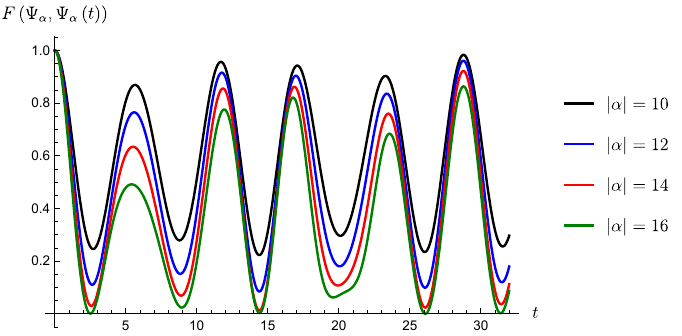}}
\subfigure[]{\includegraphics[width=0.45\textwidth]{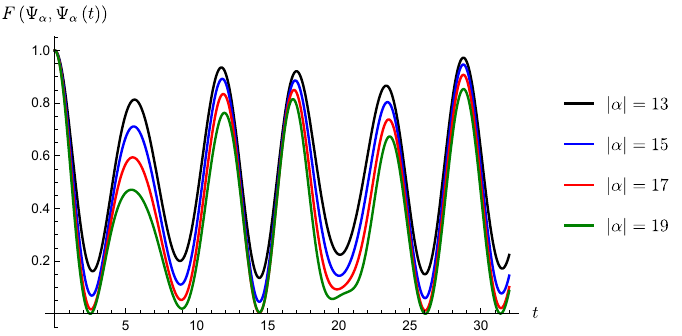}}
\caption{Fidelity plots for the BGCS of Equation \eqref{eq:56} and their time evolution for different values of the norm of the complex number $\alpha$ and (a) $p=\frac{5}{4}$, (b) $p=6$, (c) $p=18$ and (d) $p=21$. The parameters have been taken as $v_x=0.86$, $v_y=0.69$, $v_t=0.32$, $\mathcal{B}_0=1$ and $k=0$.}
\label{Figure4}
\end{center}
\end{figure}

\begin{figure}[h!] 
\begin{center}
\subfigure[]{\includegraphics[width=0.35\textwidth]{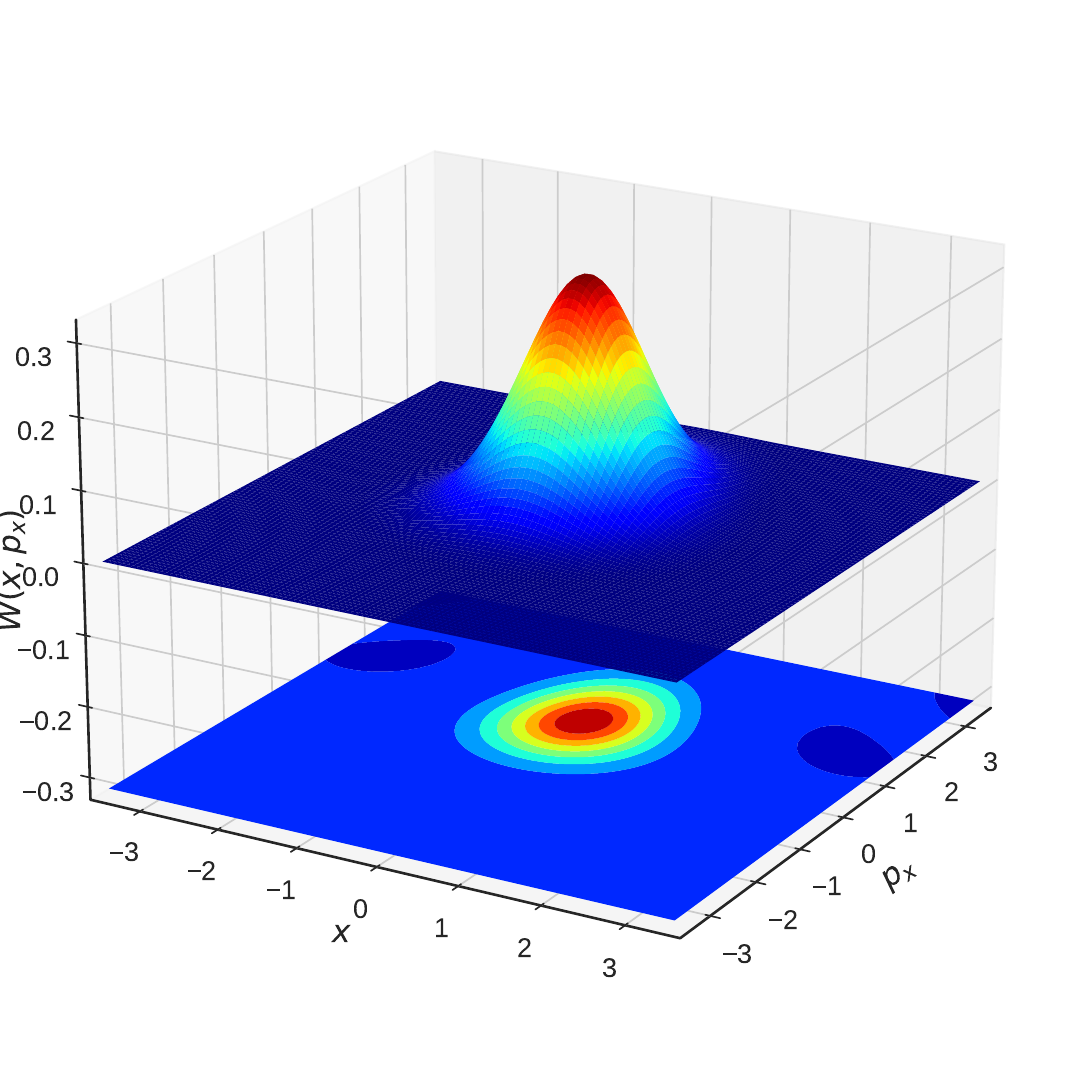}}\hspace{2cm}
\subfigure[]{\includegraphics[width=0.35\textwidth]{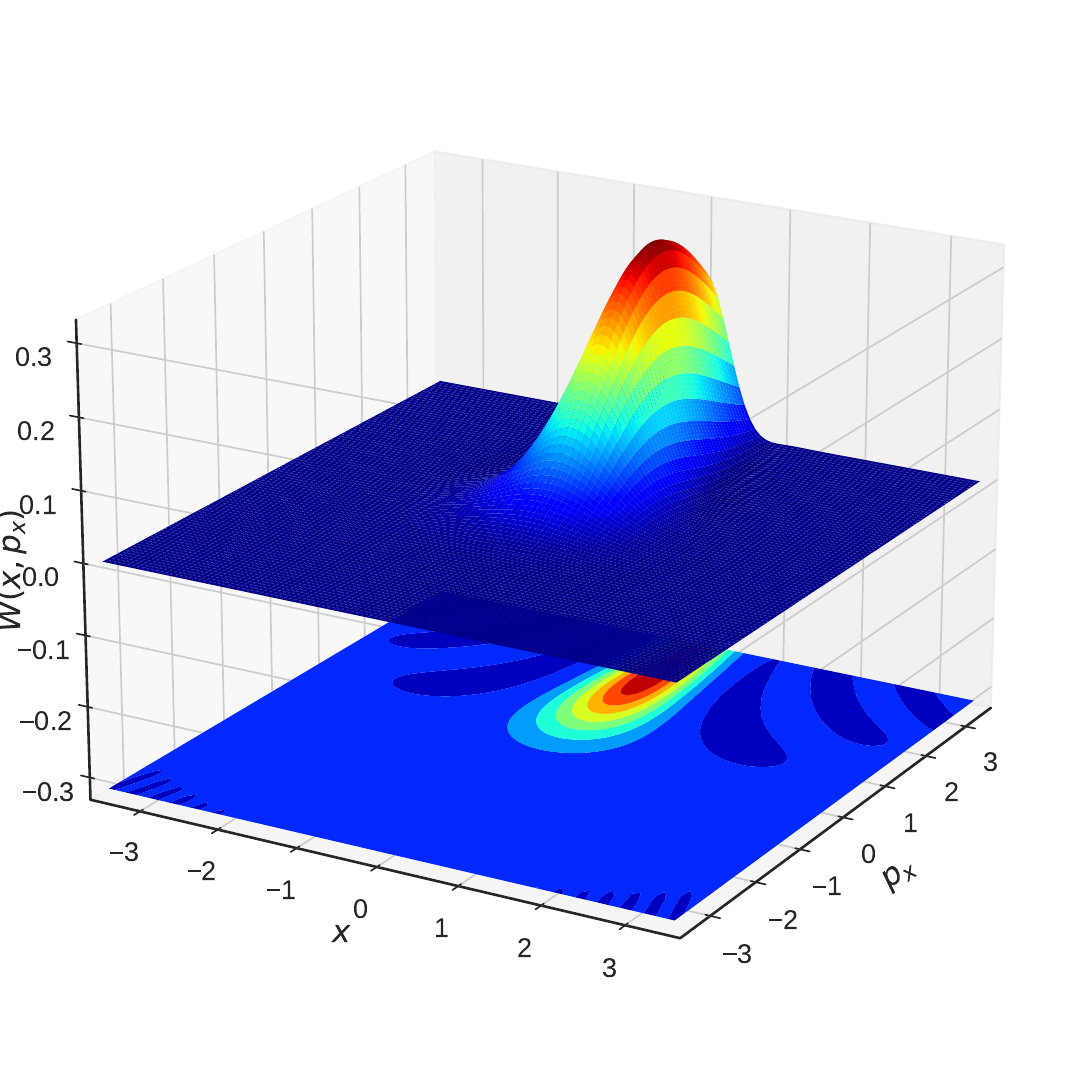}}
\subfigure[]{\includegraphics[width=0.35\textwidth]{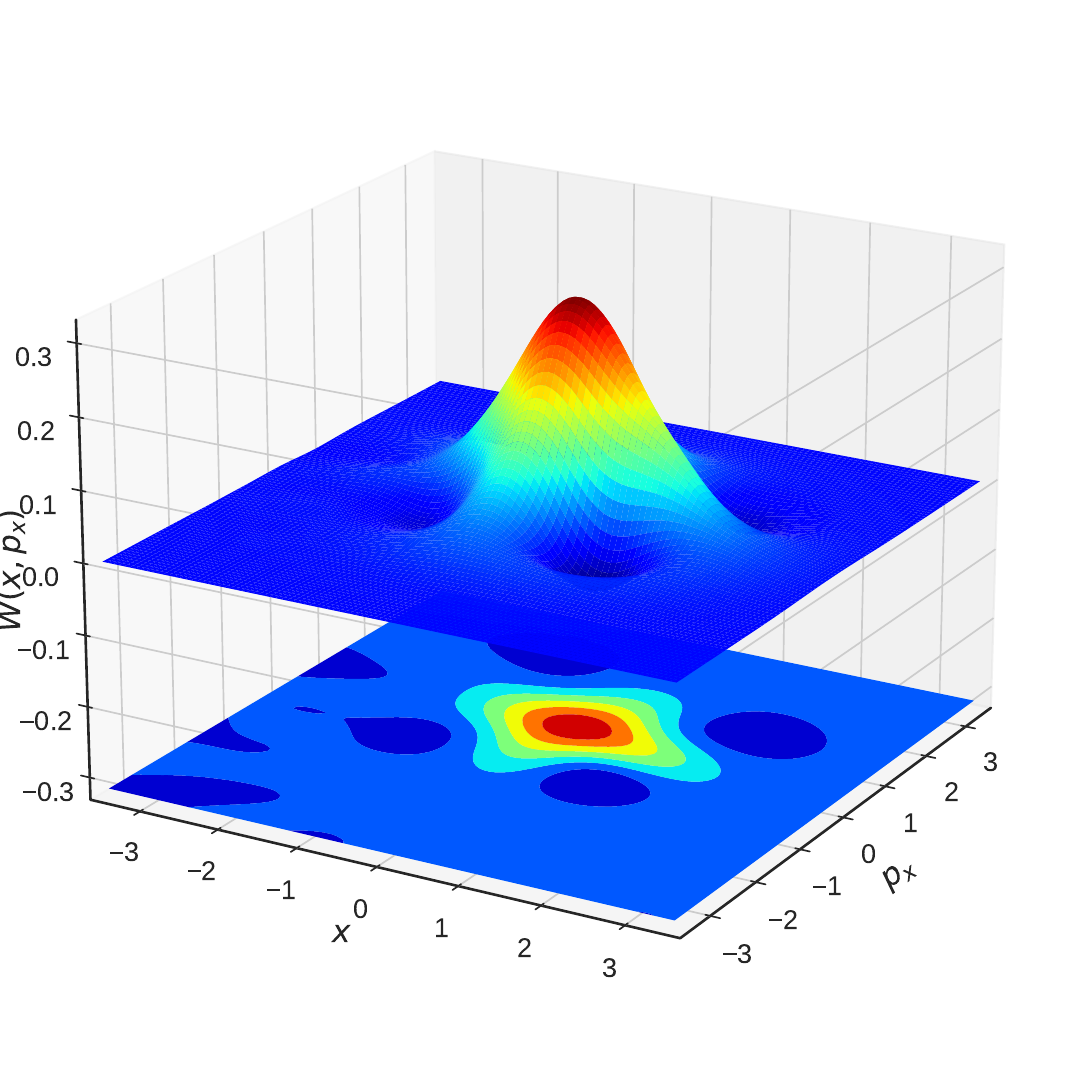}}\hspace{2cm}
\subfigure[]{\includegraphics[width=0.35\textwidth]{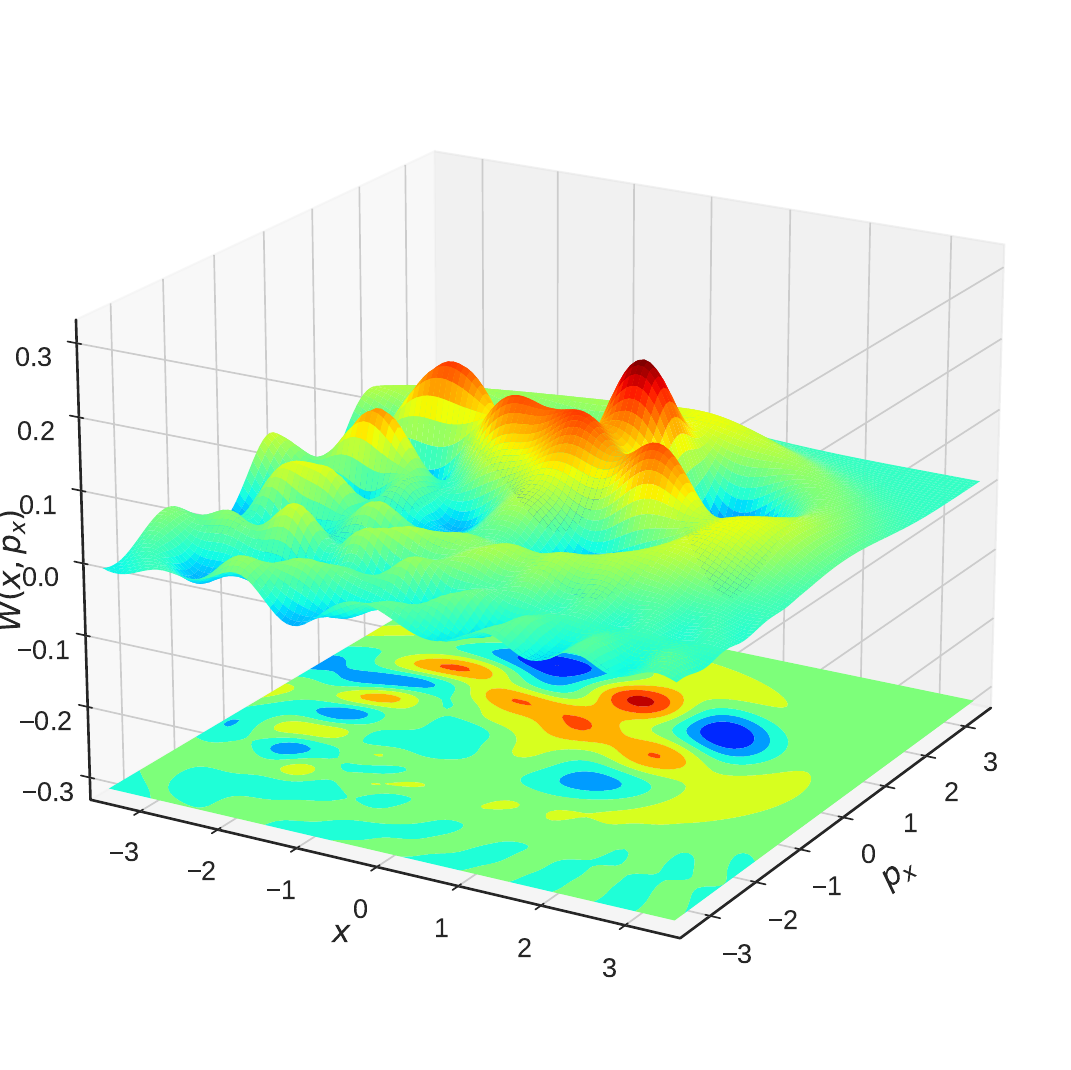}}
\subfigure[]{\includegraphics[width=0.35\textwidth]{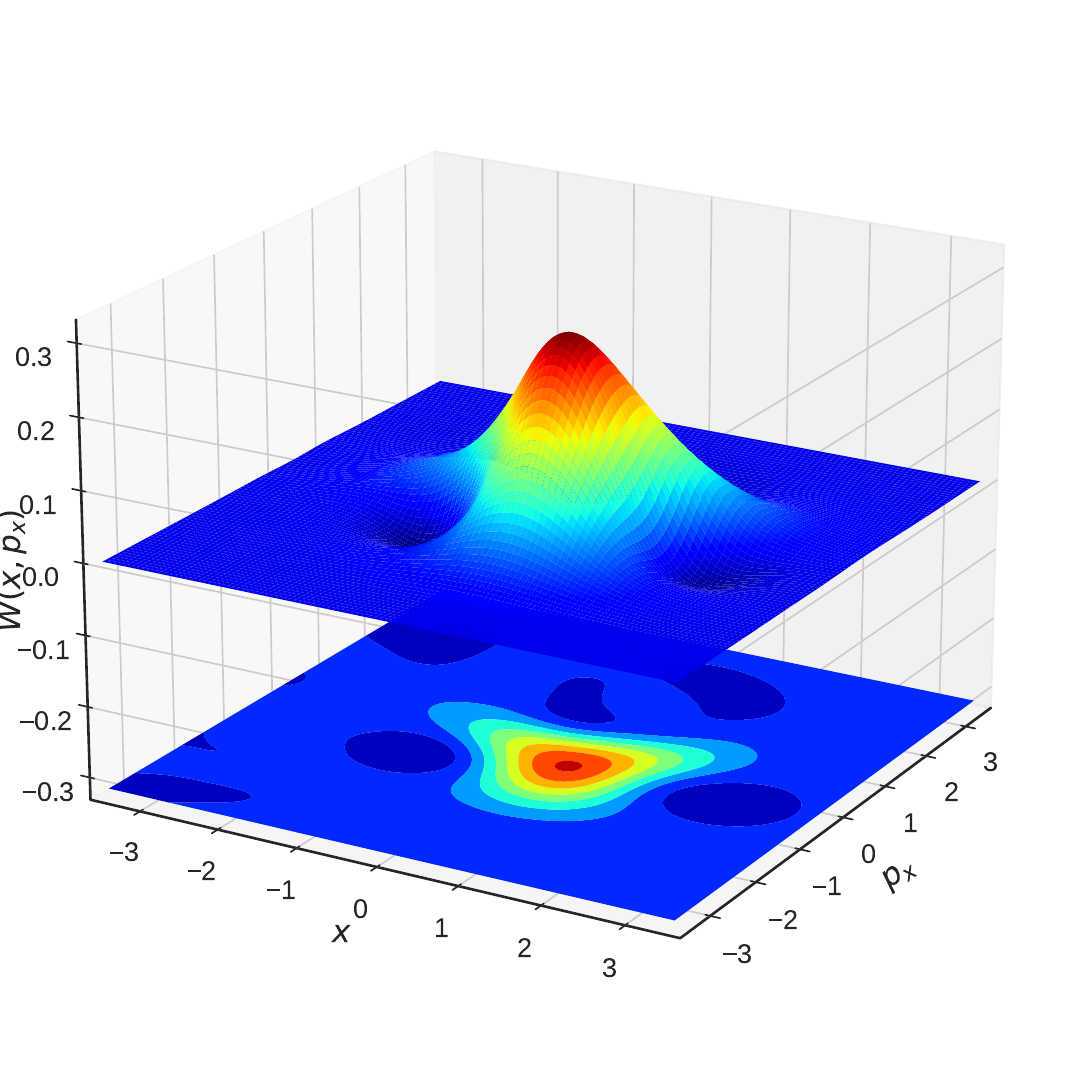}}\hspace{2cm}
\subfigure[]{\includegraphics[width=0.35\textwidth]{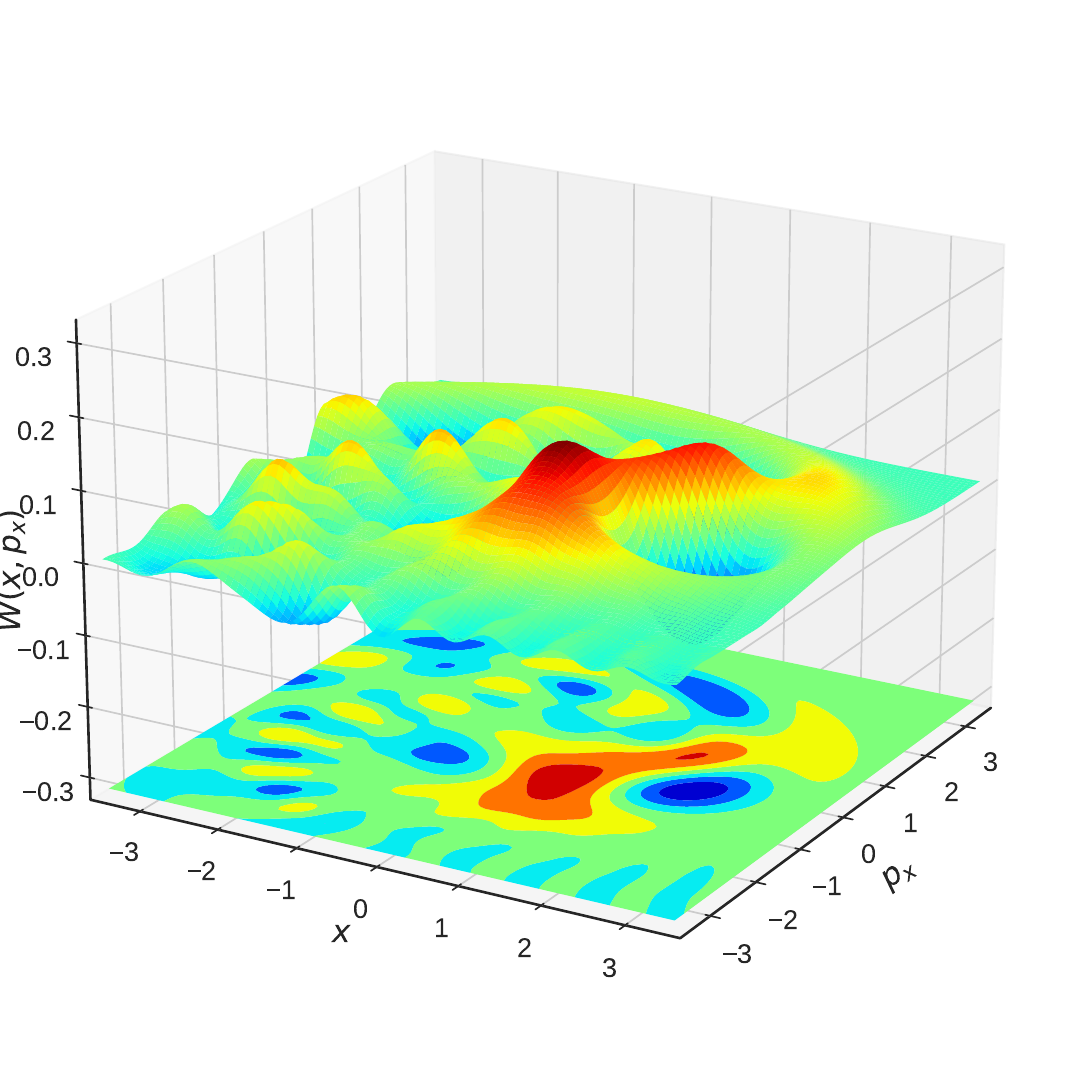}}
\caption{Plots of the trace of the Wigner function given in Equation \eqref{eq:58} with $p=21$ and $\kappa=+1$. The parameter $\alpha$ has been taken as (left column) $\alpha
=13\,{\rm e}^{i\pi/2}$ and (right column) $\alpha
=19\,{\rm e}^{i\pi/2}$ for (a, b) $t=0$, (c, d) $t=28.8$, (e, f) $t=31.2$. The parameters have been taken as $v_x=0.86$, $v_y=0.69$, $v_t=0.32$, $\mathcal{B}_0=1$ and $k=0$.}
\label{Figure5}
\end{center}
\end{figure}

Let $f_n=2^{\delta_{n,1}-2}\frac{p^2}{n}$, with $p$ any positive real number. It follows that
\begin{equation}
\left[f_n\right]!=\frac{2^{1-\delta_{n,0}}p^{2n}}{2^{2n}n!},\quad
n=0,1,...
\label{eq:55}
\end{equation}
Since the radius of convergence is $r=p$, the BGCS will converge if $|\alpha|<p$. Then, the normalization constant $\mathcal{C}_{\alpha}$ simplifies to $\mathcal{C}_{\alpha}=\frac{\sqrt{p^2-|\alpha|^2}}{p}$. In this way, the Barut-Girardello coherent state of Equation \eqref{eq:38} can be written as:
\begin{equation}
\Psi_{\alpha}(x,y)=\frac{\sqrt{p^2-|\alpha|^2}}{p}\sum_{n=0}^{\infty}
\frac{\alpha^{n}}{p^n}\Psi_n(x,y).
\label{eq:56}
\end{equation}
In such a way that the fidelity between this state and the one corresponding to its time evolution is determined by
\begin{equation}
F(\Psi_{\alpha},\Psi_{\alpha}(t))=\frac{\left(p^2-|\alpha|^2\right)^2}{p^4}
\sum_{n,m=0}^{\infty}\left(\frac{|\alpha|}{p}\right)^{2(n+m)}\cos
\left[\left(\sqrt{n}-\sqrt{m}\right)\sqrt{\omega}v_x t\right],
\label{eq:57}
\end{equation}
and the corresponding trace of 
the matrix Wigner function is
\begin{equation}
W_{\alpha}(\mathbf{r}, \mathbf{p},t)=\delta(p_y-k)\times \frac{\left(p^2-|\alpha|^2\right)}{2\pi p^2} \sum_{n,m=0}^{\infty} 
\left(\frac{|\alpha|}{p}\right)^{n+m}
\frac{ {\rm exp}\left[i(n-m)\theta+i(\sqrt{m}-\sqrt{n})\kappa v_x\sqrt{\omega}t\right]}{\sqrt{2^{2-\delta_{n0}-\delta_{m0}}}}{\rm Tr}\left[\mathcal{M}_{n,m}(x,p_x)\right].
\label{eq:58}
\end{equation}
From the previous expressions, the following can be stated: 1) Unlike the previous case, the BGCS of Equation \eqref{eq:56} do not maintain any algebraic resemblance to the standard coherent states of the harmonic oscillator. However, it is possible to identify the distribution followed by the Fock states, which is given by $P(n)=\vert a_{n}\vert^2=|\alpha|^{2n}\frac{p^2-|\alpha|^{2}}{p^{2(n+1)}}$ and 2) as in the previous case, both fidelity and the Wigner function depend on the choice of different parameters. However, here the role of $p$ plays a more significant role, as it not only determines the behavior of these functions but also determines (restricts or expands) the convergence radius of the coherent states family. Figures \ref{Figure4} and \ref{Figure5} show the plots of these functions, as well as the effect that the parameter $p$ has on them.
\subsubsection{Discussion}\label{3.3.3}
As shown in Figure \ref{Figure2} for the case $f_{n}=2^{\delta_{n,1}-2}p^{2}$, the fidelity for the BGCS in Equation \eqref{eq:52} is affected by the parameter $p$. For a $p$-value less than one (see Figures \ref{Figure2}(a) and (b)), the fidelity never reaches the value of unity for any $t$ in an interval $I$ (in this case $I$ was taken as $[0, 100]$). However, as $p$ increases, the function $F(\Psi_{\alpha},\Psi_{\alpha}(t))$ shows many oscillations in the interval $I$ whose amplitudes tend to the unity for certain values of $t$ (see Figures \ref{Figure2}(c) and (d)). This behavior is also identified in the time evolution of the Wigner function in Equation \eqref{eq:54}, as shown in Figure \ref{Figure3}. The Wigner function of the CS in Equation \eqref{eq:52} with $p=1$ does not keep its initial shape for large values of $t$, in contrast with those with $p=2$, for which the function $W_{a}(\mathbf{r}, \mathbf{p},t)$ does not modify a lot its initial form as it evolves in time.

On the other hand, for the case $f_{n}=2^{\delta_{n,1}-2}\frac{p^{2}}{n}$, where $p$ controls the radius of convergence of the corresponding states, the fidelity shows a well-defined behavior oscillatory as the parameter $p$ increases (see Figure \ref{Figure4}), which contrasts with the previous case. Figure \ref{Figure4} also shows that the fidelity for the BGCS with eigenvalue $\alpha$ such that $\frac{\vert\alpha\vert}{p}$ is near to zero, remains above for those states for which $\frac{\vert\alpha\vert}{p}\rightarrow1$. This behavior is found in the time evolution of the function $W_{a}(\mathbf{r}, \mathbf{p},t)$ as follows: by fixing the value of $p$, the Wigner functions for different values of $\vert\alpha\vert$ evolve preserving their initial shapes for a long time if the condition $0\leq\frac{\vert\alpha\vert}{p}\ll1$ is fulfilled. In contrast, as $\frac{\vert\alpha\vert}{p}\rightarrow1$, the corresponding Wigner functions do not preserve their quasi-Gaussian initial shape and take negative values as their evolve (see Figure \ref{Figure5}), which indicates the BGCS with $\vert\alpha\vert\approx p$ do not saturate the Heisenberg uncertainty relation for any $t$.

In addition, as Equations \eqref{eq:54} and \eqref{eq:58} show, the time evolution of the Wigner function for the BGCS presented here depends on the band index $\kappa$. For $\kappa=-1$, it is expected that the functions $W_{a}(\mathbf{r}, \mathbf{p},t)$ evolve in a counterclockwise direction with an angular frequency equal to $v_{x}\sqrt{\omega}$. However, we can not guarantee that, under the same initial conditions, the Wigner functions associated with BGCS with band index $\kappa=+1$ and $\kappa=-1$, respectively, are in the same place in the phase space at the same time, namely $\tau$, since the CS in Equations \eqref{eq:52} and \eqref{eq:56} do not have a well-defined period.
\section{Conclusions}\label{4}
In this work, the formalism of SUSY QM has been successfully applied to the eigenvalue problem inherent to Dirac materials, which exhibit tilted cones and are influenced by magnetic and electric fields, contrasting with previous studies \cite{DiazBautista2020, diaz2022time, betancur2021, Mojica-Zárate2024}. This formalism can be applied in cases where fields counteract the effect of the term associated with the tilt of the cones in the Hamiltonian. With the intention of approaching a semiclassical study of the problem using coherent states, a family of annihilation operators $A^-$ has been constructed defined through the typical entanglement operators of SUSY QM and characterized by the function $f$. Subsequently, the Barut-Girardello coherent states $\Psi_{\alpha}(x,y)$ were constructed as eigenstates of the annihilation operator and through a representation in the eigenfunction basis of the Hamiltonian.

Additionally, two distinct families were chosen, characterized by the selection of $f_n$ and a new parameter called $p$. The parameter $p$ plays an important role in this construction of coherent states, as depending on the choice of $f_n$, it allows us to modify the region in the complex plane where the Barut-Girardello coherent state family converges, similar to what happens with the second family in the examples (see subsubsection \ref{3.3.2}). But not only that, this parameter also enables us to manipulate the fidelity between the initial and evolved states, increasing or decreasing its value for a specific time, reducing the temporal gap between consecutive consecutive local maxima, which can be called pseudo-periods of evolution \cite{FernándezC.2022}, among other effects. These characteristics allow us to create families of coherent states that can exhibit greater (less) temporal stability, akin (in contrast) to what occurs with a coherent state of the harmonic oscillator, where the wave packet always maintains its shape. Last but not least, the study of the Wigner function allows us to analyze the behavior of the distribution of the BGCS in phase space. In general, for those BGCS as in Equation \eqref{eq:40} it can be stated that this distribution is independent of the valley index  $\nu$ and dependent on the band index $\kappa$. Furthermore, the latter acts as a time reversal transformation, and the system does not exhibit time reversal symmetry, as fidelity does not show well-defined periods of evolution.

Finally, it is important to mention that the parameter $p$ also plays a prominent role in the Wigner function. Its value allows us to modify the shape of the distribution for a defined moment in time, even allowing it to transition from a Gaussian-like or at least positive definite form to taking negative values, which indicates an increase in their quantum nature. Therefore, we can infer that for different values of $p$ and any time $t$, the Heisenberg uncertainty principle is not always saturated. In this regard, it would not be correct to say that the Barut-Girardello coherent states developed in this work are the most classical states possible. Despite this, the implementation of these states can be very useful, as many of their properties can be easily shaped by the correct choice of the function $f_n$ and the control parameter $p$.\\
\\

\noindent{\bf Acknowledgments.} The authors acknowledge financial support from CONAHCYT Project \\ FORDECYT-PRONACES/61533/2020. DOC acknowledges the support of CONAHCYT through the postdoctoral fellowship with the CVU number 712322. EDB also acknowledges the SIP-IPN research grant 20242347.
\bibliographystyle{ieeetr}
\bibliography{biblio}

\begin{thebibliography}{10}

\bibitem{Zhao2018}
Y.~Zhao, X.~Li, J.~Liu, C.~Zhang, and Q.~Wang, ``{A New Anisotropic Dirac Cone Material: A B2S Honeycomb Monolayer},'' {\em J. Phys. Chem. Lett.}, vol.~9, no.~7, pp.~1815--1820, 2018.

\bibitem{Feng20161}
B.~Feng, J.~Zhang, R.-Y. Liu, T.~Iimori, C.~Lian, H.~Li, L.~Chen, K.~Wu, S.~Meng, F.~Komori, and I.~Matsuda, ``Direct evidence of metallic bands in a monolayer boron sheet,'' {\em Phys. Rev. {\rm B}}, vol.~94, p.~041408, Jul 2016.

\bibitem{Mannix2015}
A.~J. Mannix, X.-F. Zhou, B.~Kiraly, J.~D. Wood, D.~Alducin, B.~D. Myers, X.~Liu, B.~L. Fisher, U.~Santiago, J.~R. Guest, M.~J. Yacaman, A.~Ponce, A.~R. Oganov, M.~C. Hersam, and N.~P. Guisinger, ``{Synthesis of borophenes: Anisotropic, two-dimensional boron polymorphs},'' {\em Science}, vol.~350, p.~1513, dec 2015.

\bibitem{LopezBezanilla2016}
A.~Lopez-Bezanilla and P.~B. Littlewood, ``{Electronic properties of $8\text{\ensuremath{-}}\mathit{Pmmn}$ borophene},'' {\em Phys. Rev. {\rm B}}, vol.~93, p.~241405, jun 2016.

\bibitem{Goerbig2009}
M.~O. Goerbig, J.-N. Fuchs, G.~Montambaux, and F.~Pi{\'{e}}chon, ``{{Electric-field{\textendash}induced lifting of the valley degeneracy in $\alpha$-({BEDT}-{TTF})$_2$I$_3$ Dirac-like Landau levels}},'' {\em EPL}, vol.~85, p.~57005, mar 2009.

\bibitem{katsnelson_2012}
M.~I. Katsnelson, {\em Graphene: Carbon in Two Dimensions}.
\newblock Cambridge: Cambridge University Press, 2012.

\bibitem{Cheng2017}
T.~Cheng, H.~Lang, Z.~Li, Z.~Liu, and Z.~Liu, ``{Anisotropic carrier mobility in two-dimensional materials with tilted Dirac cones: theory and application},'' {\em Phys. Chem. Chem. Phys.}, vol.~19, pp.~23942--23950, 2017.

\bibitem{Schaibley2016}
J.~R. Schaibley, H.~Yu, G.~Clark, P.~Rivera, J.~S. Ross, K.~L. Seyler, W.~Yao, and X.~Xu, ``{Valleytronics in 2D materials},'' {\em Nature Reviews Materials}, vol.~1, no.~11, p.~16055, 2016.

\bibitem{Ang2017}
Y.~S. Ang, S.~A. Yang, C.~Zhang, Z.~Ma, and L.~K. Ang, ``{Valleytronics in merging Dirac cones: All-electric-controlled valley filter, valve, and universal reversible logic gate},'' {\em Phys. Rev. {\rm B}}, vol.~96, p.~245410, Dec 2017.

\bibitem{Mrudul21}
M.~S. Mrudul, \'{A}lvaro Jim\'{e}nez-Gal\'{a}n, M.~Ivanov, and G.~Dixit, ``Light-induced valleytronics in pristine graphene,'' {\em Optica}, vol.~8, pp.~422--427, Mar 2021.

\bibitem{DiazBautista2020}
E.~D{\'{\i}}az-Bautista and Y.~Betancur-Ocampo, ``{Phase-space representation of Landau and electron coherent states for uniaxially strained graphene},'' {\em Phys. Rev. {\rm B}}, vol.~101, p.~125402, mar 2020.

\bibitem{diaz2022time}
E.~D{\'\i}az-Bautista, ``About the time evolution of coherent electron states in monolayers of boron allotropes,'' {\em Acta Polytech.}, vol.~62, no.~1, pp.~38--49, 2022.

\bibitem{betancur2021}
Y.~Betancur-Ocampo, E.~D\'{\i}az-Bautista, and T.~Stegmann, ``{Valley-dependent time evolution of coherent electron states in tilted anisotropic Dirac materials},'' {\em Phys. Rev. {\rm B}}, vol.~105, p.~045401, Jan 2022.

\bibitem{Kuru2009}
\c{S}. Kuru, J.~Negro, and L.~M. Nieto, ``{Exact analytic solutions for a Dirac electron moving in graphene under magnetic fields},'' {\em J. Phys. Condens. Matter}, vol.~21, p.~455305, oct 2009.

\bibitem{Midya_2014}
B.~Midya and D.~J. Fernández, ``Dirac electron in graphene under supersymmetry generated magnetic fields,'' {\em J. Phys. A Math. Theor.}, vol.~47, p.~285302, jun 2014.

\bibitem{Fernández_C_2020}
D.~J. Fern{\'{a}}ndez, J.~D. García-Muñoz, and D.~O-Campa, ``Electron in bilayer graphene with magnetic fields leading to shape invariant potentials,'' {\em J. Phys. A Math. Theor.}, vol.~53, p.~435202, oct 2020.

\bibitem{Fernández_C_2021}
D.~J. Fern{\'{a}}ndez, J.~D. García-Muñoz, and D.~O-Campa, ``Bilayer graphene in magnetic fields generated by supersymmetry,'' {\em J. Phys. A Math. Theor.}, vol.~54, p.~245302, may 2021.

\bibitem{FernándezC.2022}
D.~J. Fern{\'a}ndez~C. and D.~O-Campa, ``Graphene generalized coherent states,'' {\em Eur. Phys. J. Plus}, vol.~137, p.~1012, Sep 2022.

\bibitem{Wigner1932}
E.~Wigner, ``On the quantum correction for thermodynamic equilibrium,'' {\em Phys. Rev.}, vol.~40, p.~749, jun 1932.

\bibitem{Case2008}
W.~B. Case, ``{Wigner functions and Weyl transforms for pedestrians},'' {\em Am. J. Phys}, vol.~76, p.~937, oct 2008.

\bibitem{1977a}
M.~V. Berry, ``{Semi-classical mechanics in phase space: A study of Wigner's function},'' {\em Philosophical Transactions of the Royal Society of London. Series A, Mathematical and Physical Sciences}, vol.~287, p.~237, oct 1977.

\bibitem{Kenfack2004}
A.~Kenfack and K.~Zyczkowski, ``{Negativity of the Wigner function as an indicator of non-classicality},'' {\em J. Opt. B: Quantum Semiclassical Opt.}, vol.~6, p.~396, aug 2004.

\bibitem{Smithey1993}
D.~T. Smithey, M.~Beck, M.~G. Raymer, and A.~Faridani, ``{Measurement of the Wigner distribution and the density matrix of a light mode using optical homodyne tomography: Application to squeezed states and the vacuum},'' {\em Phys. Rev. Lett.}, vol.~70, p.~1244, mar 1993.

\bibitem{Hillery1984}
M.~Hillery, R.~O'Connell, M.~Scully, and E.~Wigner, ``Distribution functions in physics: Fundamentals,'' {\em Phys. Rep.}, vol.~106, p.~121, apr 1984.

\bibitem{Zachos2005}
C.~K. Zachos, D.~B. Fairlie, and T.~L. Curtright, {\em Quantum Mechanics in Phase Space}.
\newblock World Scientific, 2005.

\bibitem{BAYEN1978I}
F.~Bayen, M.~Flato, C.~Fronsdal, A.~Lichnerowicz, and D.~Sternheimer, ``{Deformation theory and quantization. I. Deformations of symplectic structures},'' {\em Annals of Physics}, vol.~111, no.~1, pp.~61--110, 1978.

\bibitem{BAYEN1978II}
F.~Bayen, M.~Flato, C.~Fronsdal, A.~Lichnerowicz, and D.~Sternheimer, ``{Deformation theory and quantization. II. Physical applications},'' {\em Annals of Physics}, vol.~111, no.~1, pp.~111--151, 1978.

\bibitem{Bordemann_2008}
M.~Bordemann, ``Deformation quantization: a survey,'' {\em Journal of Physics: Conference Series}, vol.~103, p.~012002, feb 2008.

\bibitem{Weinbub2018}
J.~Weinbub and D.~K. Ferry, ``{Recent advances in Wigner function approaches},'' {\em Appl. Phys. Rev.}, vol.~5, p.~041104, dec 2018.

\bibitem{Weinbub_2022}
J.~Weinbub and R.~Kosik, ``Computational perspective on recent advances in quantum electronics: from electron quantum optics to nanoelectronic devices and systems,'' {\em Journal of Physics: Condensed Matter}, vol.~34, p.~163001, feb 2022.

\bibitem{Leibfried1998}
D.~Leibfried, T.~Pfau, and C.~Monroe, ``{Shadows and Mirrors: Reconstructing Quantum States of Atom Motion},'' {\em Physics Today}, vol.~51, pp.~22--28, 04 1998.

\bibitem{Andrea_Bertoni_1999}
A.~Bertoni, P.~Bordone, R.~Brunetti, and C.~Jacoboni, ``{The Wigner function for electron transport in mesoscopic systems},'' {\em Journal of Physics: Condensed Matter}, vol.~11, p.~5999, aug 1999.

\bibitem{Carlo_Jacoboni_2004}
C.~Jacoboni and P.~Bordone, ``{The Wigner-function approach to non-equilibrium electron transport},'' {\em Reports on Progress in Physics}, vol.~67, p.~1033, jun 2004.

\bibitem{Jullien2014}
T.~Jullien, P.~Roulleau, B.~Roche, A.~Cavanna, Y.~Jin, and D.~C. Glattli, ``Quantum tomography of an electron,'' {\em Nature}, vol.~514, p.~603, oct 2014.

\bibitem{Bisognin2019}
R.~Bisognin, A.~Marguerite, B.~Roussel, M.~Kumar, C.~Cabart, C.~Chapdelaine, A.~Mohammad-Djafari, J.-M. Berroir, E.~Bocquillon, B.~Plaçais, A.~Cavanna, U.~Gennser, Y.~Jin, P.~Degiovanni, and G.~Fève, ``Quantum tomography of electrical currents,'' {\em Nature Communications}, vol.~10, p.~3379, 2019.

\bibitem{Fletcher2019}
J.~D. Fletcher, N.~Johnson, E.~Locane, P.~See, J.~P. Griffiths, I.~Farrer, D.~A. Ritchie, P.~W. Brouwer, V.~Kashcheyevs, and M.~Kataoka, ``Continuous-variable tomography of solitary electrons,'' {\em Nature Communications}, vol.~10, p.~5298, 2019.

\bibitem{Roussel2021}
B.~Roussel, C.~Cabart, G.~F\`eve, and P.~Degiovanni, ``Processing quantum signals carried by electrical currents,'' {\em PRX Quantum}, vol.~2, p.~020314, May 2021.

\bibitem{weyl1931theory}
H.~Weyl and H.~Robertson, {\em The Theory of Groups and Quantum Mechanics}.
\newblock Dover books on advanced mathematics, Dover, 1931.

\bibitem{Mojica-Zárate2024}
J.~A. Mojica-Z{\'a}rate, D.~O-Campa, and E.~D{\'i}az-Bautista, ``An algorithm for exact analytical solutions for tilted anisotropic dirac materials,'' {\em Eur. Phys. J. Plus}, vol.~139, p.~272, Mar 2024.

\end{thebibliography}
------------------------
\end{document}